%% file: ms.tex
\documentclass{sig-alternate}

\usepackage{algorithm}
\usepackage{algpseudocode}
\newcommand\Subroutine[1]{%
  \vspace*{-.7\baselineskip}\Statex\hspace*{\dimexpr-\algorithmicindent-2pt\relax}\rule{\columnwidth}{0.4pt}%
  \Statex\hspace*{-\algorithmicindent}\textbf{Subroutine - }{#1}%
  \vspace*{-.7\baselineskip}\Statex\hspace*{\dimexpr-\algorithmicindent-2pt\relax}\rule{\columnwidth}{0.4pt}%
}
\algblockdefx{ParFor}{EndParFor}[1]{\textbf{parallel for} #1}{\textbf{end parallel for}}
\usepackage{array}
\usepackage{booktabs}
\usepackage{graphicx}
\usepackage{grffile}            
\usepackage{multirow}
\usepackage[caption=false,font=footnotesize]{subfig}
\usepackage{tikz}
\usepackage{url}
\usepackage{xcolor}

\graphicspath{{./figures/}} 

\begin{document}


\input{custom_commands}





%

\title{Field of Groves: An Energy-Efficient Random Forest}
%
%
%
%
%

\numberofauthors{1} 
%
\author{
\alignauthor
Zafar Takhirov,
Joseph Wang,
Marcia S. Louis\\
Venkatesh Saligrama,
Ajay Joshi
%
%
\and
\affaddr \{zafar, joewang, marcia93, srv, joshi\}@bu.edu\\
\affaddr{Boston University, Boston, MA}
}

\maketitle
\input{abstract}



%
%

%
%


\setlength{\textfloatsep}{10pt} 
\input{introduction}
\input{related-work}
\input{proposed-approach}

\input{evaluation}

\input{conclusion}


%

\bibliographystyle{abbrv}
\bibliography{main}  
%
%
\end{document}

%% file: custom_commands.tex
\newcommand{\ignore}[1]{}
\newcommand{\fixme}[1]{{\color{red}\bfseries [ FIXME: #1 ]}}
\newcommand{\ajay}[1]{{\color{blue}\bfseries [ AJAY: #1 ]}}
\newcommand{\zafar}[1]{{\color{green!70!black}\itshape [ ZAFAR: #1 ]}}
\newcommand{\joe}[1]{{\color{purple}\bfseries [ Joe: #1 ]}}

\newcommand{\boxaround}[2][\columnwidth]{}
\DeclareRobustCommand\circled[1]{\tikz[baseline=(char.base)]{
    \node[shape=circle,draw,inner sep=2pt] (char) {#1};}}
\newcommand{\argmin}{\operatornamewithlimits{argmin}}
\newcommand{\argmax}{\operatornamewithlimits{argmax}}

\newcommand*{\al}[2]{\multicolumn{1}{#1}{#2}}
\newcommand*{\vertrows}[2]{\multirow{#1}{*}{\rotatebox[origin=c]{90}{#2}}}
\newcommand{\minitab}[2][l]{\begin{tabular}{#1}#2\end{tabular}}
\newcommand*\etal{{\it et al.}}
\providecommand{\e}[1]{\ensuremath{\times 10^{#1}}} 

\newcommand{\missingcitation}{[{\color{cyan}\textbf{??}}]~}


%% file: abstract.tex
\begin{abstract}
Machine Learning (ML) algorithms, like Convolutional Neural Networks (CNN), Support Vector Machines (SVM), etc. have become widespread and can achieve high statistical performance. However their accuracy decreases significantly in energy-constrained mobile and embedded systems space, where all computations need to be completed under a tight energy budget. In this work, we present a field of groves (FoG) implementation of random forests (RF) that achieves an accuracy comparable to CNNs and SVMs under tight energy budgets. \ignore{The proposed field of groves implementation is scalable and can be dynamically tuned to meet varying energy budgets or accuracy demands.}
Evaluation of the FoG
shows that at comparable accuracy it consumes $\approx$1.48$\times$,
$\approx$24$\times$, $\approx$2.5$\times$, and $\approx$34.7$\times$ lower energy per classification
compared to conventional RF, SVM$_{RBF}$, MLP, and CNN, respectively. FoG is $\approx$6.5$\times$ less energy efficient than SVM$_{LR}$, but achieves 18\% higher accuracy on average across all considered datasets.
\ignore{
In the Big Data age, the volume of data that is continually generated is
increasing, and extracting knowledge from this volume of data is increasingly
challenging. Machine Learning (ML) approaches are commonly deployed today for
knowledge extraction. Conventional (ML) algorithms, like convolutional neural
networks (CNNs), multi-layer perceptron (MLP), support vector machines (SVMs),
etc. are \ignore{pervasive}prevalent and can achieve high statistical performance, but their
accuracy decreases significantly in energy-constrained mobile systems space,
where all computations need to be completed under a tight energy budget. In this
work, we show the use of Random Forests (RFs) as an alternative to CNN, MLP and
SVM for energy-constrained environments. We present a Field of Groves
implementation of RFs that efficiently performs inference under tight energy
budgets. By examining decision confidence for each example, computational
resources are dynamically allocated to examples with the highest uncertainty,
reducing the need to expend energy on examples with low uncertainty. The
proposed field of groves implementation is scalable and can be dynamically tuned
to meet varying energy budgets or accuracy demands. We evaluate our proposed
design on several datasets using energy and accuracy metrics. Our evaluation
shows that the accuracy of the traditional RF classifier is comparable\ignore{(if not
larger)} to CNN for all datasets that we considered and at the same time RF
consumes $\approx$ 95.4\%, 91.5\%, and 99.4\% less energy than SVM$_{RBF}$, MLP
and CNN, respectively. The maximum achievable accuracy of the FoG implementation
is lower than RF and CNN by 3.2\% and 4\%, respectively, but FoG classifiers has
42\% and 99.7\% lower energy than RF and CNN, respectively.
}
  \ignore{
    In today's Big Data age the volume of data that gets generated everyday is
    increasing. Extracting knowledge from this large volume of data is becoming
    increasingly challenging. Machine Learning (ML) approaches are commonly deployed
    today for knowledge extraction. Conventional (ML) algorithms, like convolutional
    neural networks (CNNs), Support Vector Machines (SVMs), etc. are quite pervasive
    and can achieve high statistical performance, but their accuracy decreases
    significantly in energy-constrained mobile and the embedded systems space, where
    all the computations need to be completed under tight energy budget. In this
    work we show the use of Random Forests (RFs) as an alternative to CNN and SVM
    for energy-constrained environments. As part of the paper, we present a
    multi-grove implementation of RFs. \fixme{Need a sentence here about how
      multi-grove implementation uses the confidence to decide whether to use more
      groves or not, which in turn avoids the need to spend extra energy in case of
      easier data} The multi-grove implementation is scalable can be dynamically tuned
    to meet varying energy budgets or accuracy demands. We validate our proposed
    design on several datasets using energy and accuracy metrics. We also compare
    our multi-grove implementation against other ML algorithms including SVM, CNN,
    Multi-Layer Perceptron, and k-Nearest-Neighbors. We show that in
    energy-constrained environments, compared to the other ML algorithms multi-grove
    implementation consumes lower energy at comparable accuracy and has better
    accuracy at comparable energy consumption.

    \fixme{We should formulate a name for this multi-grove implementation. The name
      could be an acronym based on the title.}
  }
  \boxaround{
  {\em Skeleton of the abstract:}
  \begin{itemize}
  \item Neural nets, although highly accuracy, fail in certain conditions
  \item Budget constrained environment requires different algorithms
  \item Random forest algorithm is one of the suitable candidates
  \end{itemize}
  Three main takeaways:
  \begin{enumerate}
  \item {\bf Algorithmic} Novel implementation of random forest algorithm
  \item {\bf Architectural} Novel approach to resource sharing and allocation between computing blocks (only on ASIC/FPGA):
    \begin{enumerate}
    \item Queued parallel cache
    \item Asynchronous communication between computing units
    \item Easily scalable architecture
    \end{enumerate}
  \item {\bf Automation and Test:} Design flow + modeling (this one is similar to Brooks' latest ISCA work)
  \end{enumerate}
  }
\end{abstract}

%% file: introduction.tex
\section{Introduction}
\label{sec:introduction}
\boxaround{
\begin{itemize}
    \item {\bf Motivation and problem statement}
    \begin{enumerate}
        \item {\bf Motivation:} As the level of integration of the ML algorithms increases, the need for energy-efficient implementation becomes critical.
        \item {\bf Statement:} Being on tight energy budget, the mobile and embedded devices have different design goals. In mobile devices energy-efficiency is more important than energy itself. Energy-efficiency in mobile devices is a subject to current energy, delay, or accuracy budgets. At the same time, in embedded systems the energy-efficiency is subject to energy and accuracy.
        \item {\bf Problem:} Some algorithms, although providing great benefits in accuracy-centric environments, fail when working in energy-limited ones
        \item {\bf Solution:} Carefully pick a more energy-efficient algorithm by sacrificing some of the accuracy
        \item {\bf Problem:} Mobile and embedded systems change the operating environments depending on current battery reserves, performance requirements, accuracy goals, etc.
        \item {\bf Solution:} When designing the algorithms, design them with an ability to adapt and/or tune to different requirements
    \end{enumerate}
    \item {\bf Subsections}
    \begin{enumerate}
        \item ML Applications
        \item ML Algorithms comparison
    \end{enumerate}
\end{itemize}
}

Over the last couple of decades the consumer market has gradually moved towards
mobile computing. According to the comScore report, people of age 18-64 spend
$\approx$70\% of their digital time using a mobile device \cite{comscore:2016}.
Mobile workloads
are increasingly data intensive, and hence machine learning (ML) algorithms are
commonly used in these applications~\cite{5560598}. The data-intensive
nature of these applications does not allow us to run these applications purely
on our mobile systems. For example, applications like speech recognition,
although used quite often, are still evaluated remotely because limited energy
budgets prohibit running a powerful ML algorithm\ignore{machine} on a mobile device.
\ignore{Another example is handwriting recognition -- open source NBIS library
from NIST uses multi-layer perceptron for handwriting identification and
verification \cite{nist:nbis}. As another example, The work done by Le et al. in
collaboration with Google Inc. has shown a deep network that had 1 billion
trainable parameters \cite{ranzato:icml:2012} can have very high accuracy in the
applications like ImageNet. However, such a deep neural network is extremely
cumbersome to operate purely on a mobile system.}

Mobile systems\ignore{however,} are energy constrained, and hence while designing
machine learning architectures for mobile applications we need to manage two
conflicting requirements -- high accuracy and low energy dissipation. Over the past few years several architecture-, circuit-
and algorithm-level optimizations have been proposed for improving the power,
performance, area, and accuracy of machine learning accelerator designs
\cite{Du:2015:NAC:2830772.2830789, takhirov:islped:2016, ASTC_AAAI14}\ignore{, Gao+Koller:NIPS11,park2012fpga,DBLP:conf/icml/XuWC12,
trapeznikov:2013b, NIPS2015_5982, ASTC_AAAI14, icml2015_nan15,
maashri2012accelerating, albericiocnvlutin, han2016eie,
chen2016eyeriss, shafiee2016isaac, chi2016prime, rahman2016efficient}.
\ignore{Too
many references. We just need to list the ones that were published in DAC, DATE,
ISCA, MICRO, HPCA, ICCAD.} \ignore{Several
researchers have also shown in the past that high accuracy is not always required for
all applications, and approximations are sometimes adequate
\cite{shafique:dac:2016,
akhlaghi:date:2016}}\ignore{these works just show that sometimes approximations
are adequate, they also discuss different algortihms and approaches to
approximate evaluation}\ignore{tziantzioulis:date:2016,gupta2014nondet,sampson2015diss,Are these papers talking about how we can
tolerate approximations in applications in general or are they specifically
focusing on applications that use machine learning? -- which allows us to
tradeoff accuracy for energy efficiency when choosing the machine learning
algorithms.}

\ignore{The following argument about training time does not make sense to me.
Why are we talking about this?}\ignore{Another parameter that one might
consider when deciding on an accelerator algorithm is the training time -- the
number of minutes or hours necessary to train a model varies between algorithms
and is usually directly related to the accuracy. There has been a lot of work
that focuses on minimizing the training time \cite{7436772, 7472150, MAL-006,
Huang201532} However, machine learning algorithms that operate on a mobile or
embedded system require fast and energy-efficient run-time evaluation, and it is
more beneficial to perform the training offline, and focus on optimizing the
run-time Figures-of-Merit (FoM) \cite{takhirov:islped:2016, trapeznikov:2013b,
wang2014model}.}

\ignore{Table \ref{table:ml-comparison} shows a coarse comparison for several
different design considerations (derived from \cite{microsoft:azure}). We are
not giving all possible metrics for the ML algorithms and the valuation given in
the table is for discussion purposes only. The ``Run-Time Tunability'' is a
custom defined metric, and represents how easy it is to tune some of the
components of the algorithm to achieve better energy-efficiency (hardware
implementation only).

\begin{table}[tbh]
    \centering
    \caption{Machine Learning Classifiers Cheat Sheet -- Note that the choice of classifiers is
    determined by more than a handful of parameters. In addition to that all classifiers are
    highly dependent on the Kernel functions.
    \joe{I don't understand this table at all. Log. Reg., Perceptron, and SVM are all highly dependent on
    the Kernel functions chosen, making it hard to compare accuracy or energy efficiency. Additionally,
    training-time is not really a field of interest in this paper. I would remove this table as it's really
    easy to offend someone who loves SVM or NN with a table like this.}
    }
    \label{table:ml-comparison}
    \begin{tabular}{lcccc}
        \toprule
        ~                   & ~                     & Energy                & Run-time      & Training                      \\
        Algorithm           & Accuracy              & Efficiency            & Tunability    & Time                              \\
        \midrule
        Logistic Regression & $\times$              & \checkmark\checkmark  & --            & \checkmark\checkmark              \\
        Random Forest       & \checkmark\checkmark  & \checkmark            & \checkmark    & \checkmark   \\
        Boosted Dec. Tree   & \checkmark\checkmark  & \checkmark            & \checkmark            & \checkmark   \\
        Neural Network      & \checkmark\checkmark  & $\times$              & \checkmark    & $\times$ \\
        Averaged Perceptron & \checkmark            & $\times$              & --            & \checkmark \\
        SVM                 & --                    & --                    & --            & -- \\
        \midrule
        Proposed Approach   & \checkmark\checkmark  & \checkmark            & \checkmark\checkmark    & --   \\
        \bottomrule
    \end{tabular}
\end{table}
}
\ignore{
In this paper, we explore the use of random forest classifiers (RF) for energy-efficient
classification when operating in
}
The machine learning community has shown that we do not always
require complex classifiers such as convolutional neural networks (CNN) or kernel
support vector machines (SVM) for classifying data, and that low-complexity 
classifiers such as random forests (RFs) are an adequate substitute for applications
where high accuracy with low energy dissipation is required
\ignore{Gao+Koller:NIPS11,}\cite{icml2015_nan15}. In this paper, we propose an alternative implementation of RF classifiers. We divide the RF into groups of trees called groves for budget-constrained environments, where the
budget is accuracy, energy, delay or energy-delay product. In general, RFs are a
collection of decision trees (DTs) that independently predict the classification
result, with the final decision made by combining the decisions of
individual trees. Our approach uses the confidence of groves within the RF about their decision to optimize the resource utilization.
 In this work:\\
-- We first evaluate the use of RF algorithm as an alternative
to CNN, SVM with linear (SVM$_{LR}$)\ignore{red SVM with logistic regression is not a real thing, are you doing logistic regression or linear SVM? If doing logistic regression, just call it logistic regression (LR), otherwise call it linear SVM.} and with radial-basis function
(SVM$_{RBF}$) as the kernels, and Multi-layer Perceptron (MLP) algorithms. Our analysis shows that the RF accuracy
is comparable to the accuracy of CNN, SVM$_{LR}$, SVM$_{RBF}$ and MLP
for all evaluated datasets, and on average RF consumes $\approx$10$\times$
lower energy per classification. \ignore{Need to double check these numbers}\\
\ignore{
-- We propose a novel implementation of RF called {\bf Field of
Groves (FoG)}. In the FoG implementation, RF is subdivided into smaller
forests, which we call groves. These groves have smaller number of DTs and evaluate their decision confidence in parallel and independently
from each other. Each grove decides if more groves
need to process its current input depending on the computed confidence
and the confidence threshold. That way more computational
resources are dynamically allocated to examples with higher uncertainty,
while a smaller number of groves are allocated to the examples with lower uncertainty, thus reducing the average energy dissipation. Our evaluation
shows that at comparable accuracy FoG consumes $\approx 1.48\times$,
$\approx 24\times$, $\approx 2.5\times$, and $\approx 34.7\times$ lower energy per classification compared to
conventional RF, SVM$_{RBF}$, MLP, and CNN, respectively, while having similar
energy dissipation as SVM$_{LR}$ \fixme{make sure to double check the numbers}
}
-- We also propose a novel implementation of RF called Field of Groves (FoG).
FoG is composed of multiple groves, where every grove is a subset of
decision trees. During the evaluation period, the groves start the class
probability estimations in parallel, with every grove receiving different
inputs. If the probability threshold (confidence level) is not met, the
``partially computed'' result is issued to the next grove. That way more
computational resources are dynamically allocated to examples with higher
uncertainty, thus reducing the average cost of estimation. Our evaluation
shows that at comparable accuracy FoG consumes $\approx$1.48$\times$,
$\approx$24$\times$, $\approx$2.5$\times$, and $\approx$34.7$\times$ lower energy per classification
compared to conventional RF, SVM$_{RBF}$, MLP, and CNN, respectively, while having
similar energy dissipation as SVM$_{LR}$. \ignore{make sure to double check the
numbers}

\ignore{Section \ref{sec:proposed-approach} provides detailed description of the RF
algorithm and our FoG implementation of the RF algorithm. Section
\ref{sec:evaluation} shows the comparison of the traditional RF implementation
and the FoG implementation of RF with CNN, MLP, and SVM in terms of accuracy and
energy efficiency, followed by concluding remarks in
Section~\ref{sec:conclusion}.}

%% file: related-work.tex
\begin{figure}[htp]
	\centering
    \includegraphics[width=.65\columnwidth]{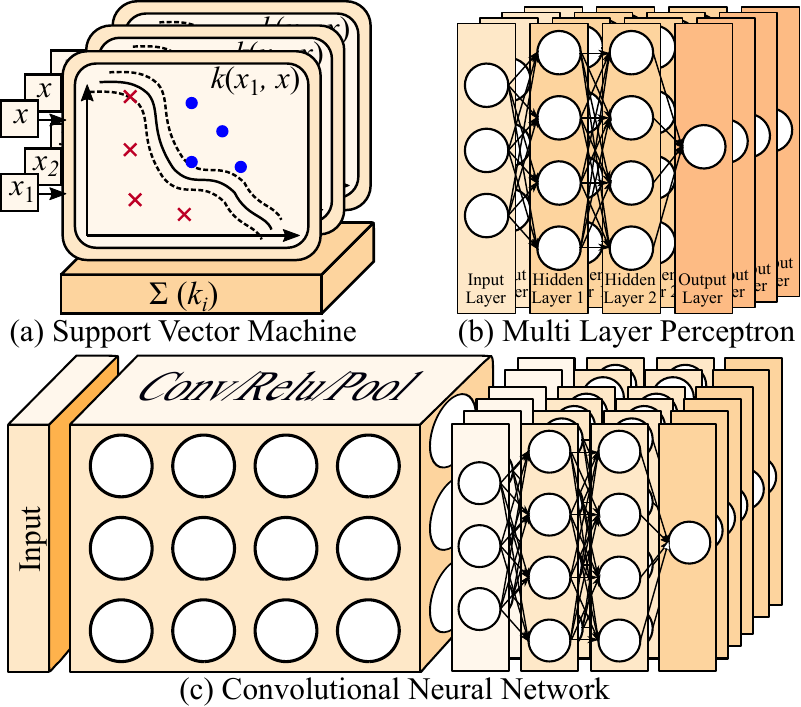}
	\caption{Logical view of SVM, MLP, and CNN.}\label{fig:logical:svm-mlp-cnn}
\end{figure}

\begin{figure}[htp]
	\centering
    \includegraphics[width=\columnwidth]{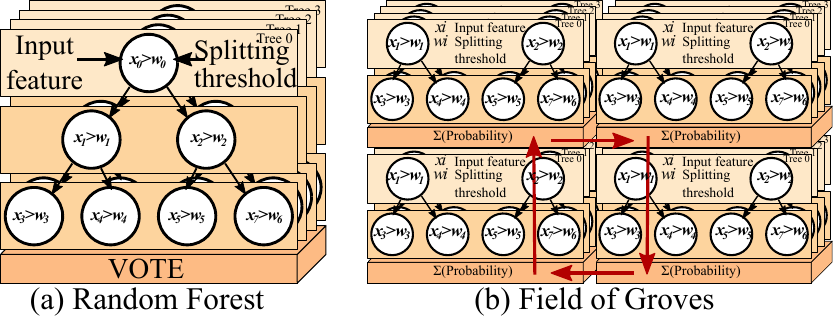}
    \caption{Logical view of RF and FoG.}\label{fig:logical:rf-fog}
\end{figure}
\section{Related Work}
\label{sec:related-work}
Algorithmically, our proposed approach is based on RF classifiers
\cite{breiman}. Traditionally, energy efficiency has not been considered when
designing random forests; however, recent work has studied learning of the RFs as a subject
to test-time constraints \cite{icml2015_nan15}. This approach centers on
reducing feature/sensor acquisition cost, however it does not address the system
energy constraints. Similar approaches to learning decision rules to minimize
error subject to a budget constraint during prediction-time have been proposed
\cite{icml2015_nan15,DBLP:conf/icml/XuWC12,
NIPS2015_5982}. Although closely related, these
approaches also ignore the energy usage and disregard computational cost, making
them of limited use in an energy constrained settings.

The state of the art machine learning techniques for image processing, video
processing, object recognition, etc. are CNN and DNN, and custom hardware
accelerator design have been proposed for the same
\ignore{chakradhar2010dynamically}\cite{park20154}.
In addition to that, a variety of techniques including stochastic computing \cite{kim2016dynamic}, dataflow
architecture \cite{chen2016eyeriss, nowatzki2015exploring},\ignore{chen2016eyeriss} data reuse
\cite{rahman2016efficient}, custom sparse matrix-vector multiplication
\cite{han2016eie} and run-time adaptivity \cite{venkataramani:2015:sce,
takhirov:islped:2016} have been adopted to achieve energy-efficient machine learning algorithm operation.

\ignore{
Accelerators aim to achieve energy efficient computation by using different
techniques, in \cite{kim2016dynamic} to reduce the area, power and energy
requirements of conventionally DNNs, they apply stochastic computing to DNNs.
The goal is achieved by removing near-zero weights, using weight scaling and
state machines for activation function with an accumulator. They also propose
early decision termination to save energy and increase decision speed. In
Cnvlutin(CNV) \cite{albericiocnvlutin} which is a value based DNN accelerator
that uses a Zero-Free Neuron Array format to store the input for each layer
which dynamically eliminates most ineffectual multiplications. The primary focus
of CNV is on the convolutional layer as it dominates execution time. The
CNV design is based on the state of the art DaDianNao
\cite{chen2014dadiannao} accelerator design with improvement in performance and
energy. Another method for energy efficiency computation is to use dataflow
architecture \cite{1253203, nowatzki2015exploring}, Eyeriss
\cite{chen2016eyeriss} is a spatial architecture based on dataflow in different
layers of CNN. The dataflow is hierarchically categorized into several levels
called row stationery(RS). In the new dataflow model, each row pair stays
stationary to the Processing Element (PE). It increases the throughput of both
convolutional and fully-connected layers and optimizes all types of data
movement in the storage hierarchy. Input-recycling Convolutional Array of
Neurons (ICAN) \cite{rahman2016efficient}, is a flexible 3D compute title that
implements the convolution operations used in DNNs. ICAN is efficient and
supports a range of convolution layer shapes. ICAN is used along with Input
Reuse Network, which is a 2D mesh-like array of registers to facilitate data
reuse. EIE \cite{han2016eie}, an efficient inference engine, operates on
compressed Deep Neural Network. It is a specialized accelerator designed with an
array of processing elements (PEs) to perform customized sparse matrix-vector
multiplication and efficiently handle weight sharing. A part of the network is
stored in SRAM of a PE, and it uses that part for its computation.

Energy efficiency can be achieved by reducing the data movement, Processing In
Memory architecture for NN, ISAAC \cite{shafiee2016isaac} is an in-situ analog
accelerator that is similar to DaDianNao \cite{chen2014dadiannao} in system
organization, as it has multiple nodes/tiles but with memristor crossbar arrays
in each tile. These crossbar stores dedicated a set of neuron and perform the
dot-product computations. The design is easy to pipeline because outputs of the
previous layer are fed to other crossbars devoted to processing the next CNN
layer, and so on. More crossbars with same weights are created to improve the
throughput of a layer. Another Memory-centric hardware design for Boltzmann
machine is proposed in \cite{bojnordi2016memristive}, it uses resistive RAM
(RRAM) crossbar for dot product computation and also stores the synaptic weights
in them. PRIME \cite{chi2016prime} is another Processing In Memory architecture
for NN that uses metal-oxide resistive random access memory (ReRAM) crossbar for
dot product computation. The ReRAM main memory is designed such that a part of
the memory is NN accelerator, and the rest is the memory. This partition of
ReRAM is dynamically reconfigured between memory ad accelerators with
architecture and software support.

Other works focus on minimizing the evaluation cost using the input properties.
Venkataramani et al. \cite{venkataramani:2015:sce} centers on using
increasingly complex classifiers on examples close to the decision boundary.
This system assumes that complex classifiers will always correctly classify
examples that are incorrectly classified by simple models, ignoring the behavior
of complex classifiers when determining the appropriate classifier for each
example. This potentially might cause an extremely hard example to go through
all classification stages, and still incorrectly identify the label, thus
wasting energy. Additionally, for this approach, trading off between
classification error and energy efficiency is difficult to adjust, particularly
when presented with multiple classification functions.

Hierarchical framework of classifiers was also investigated by Panda et al.
\cite{panda:arxiv:2015}, but the work also suffers from the lack of dynamic
trade-off adjustment. Park et al. \cite{park:icc:2015} proposed a ``big-little''
topology for neural networks. The work proposed by Takhirov et al.
\cite{takhirov:islped:2016} uses a set of classifiers and a ``chooser'' function
to dynamically choose the complexity of a classifier. They also employ a bias
variable for dynamic tuning. The main drawback of their approach is that the
classifiers have to be increasingly more complex, and thus don't have additive
properties.}

The current work proposes a novel\ignore{energy-efficient solution to budget constrained
evaluation using} implementation of random forest ensemble learning method
\ignore{In contrast to other works, the proposed approach focuses }with the
focus on energy-efficient solution
to budget constrained evaluation.

%% file: proposed-approach.tex
\section{Random Forests}
\label{sec:proposed-approach}
\input{random-forests}
\input{groves}

%% file: random-forests.tex
\subsection{Random Forests: Conventional Design}
\label{sec:random-forests}
\boxaround{
\begin{enumerate}
  \item {\bf Content}
  \begin{enumerate}
    \item Brief description of what are Random Forests. Also talk about them being comparable if the environment is energy-constrained.
    \item Description of the training procedure and hardware construction
    \item Note that the random forests are easily scalable
    \item Random forests could be implemented from any types of classifiers, but for this work we used decision trees
  \end{enumerate}
  \item {\bf Images}
  \begin{enumerate}
    \item Block diagram of a random forest implementation on hardware
  \end{enumerate}
  \item {\bf Algorithms}
  \begin{enumerate}
    \item Training + evaluation
  \end{enumerate}
\end{enumerate}
}

As mentioned in Section~\ref{sec:introduction}, we commonly use SVM as well as traditional
neural network-based algorithms like CNN or MLP for classifying data
sets with large number of features. Figure~\ref{fig:logical:svm-mlp-cnn} shows
high-level logical view of SVM, MLP, and CNN. In this section we analyze the use of RFs as compared to the popular classification algorithms.

\ignore{CNNs, which can achieve very high statistical performance, typically contain
multiple convolutional and pooling layers that are followed by a fully connected classifier like MLP. Here the convolutional and pooling layers
perform filtering, feature extraction, and reduce
dimensionality of the classification problem, which can then be easily handled
by smaller MLPs. A pure MLP, without convolutional layers,
might require tens of hidden layers and hundreds of nodes more to achieve
statistical performance similar to CNN, if at all possible \cite{Goodfellow-et-al-2016-Book}.
Classifiers like SVM use specialized functions
(kernels) to create a (many-dimensional) boundary between classes. Such
classifiers, once designed, are hard to tune at run time, and their operating
points are usually defined at design time. Although it is possible to change
the ``support vectors'' in the SVM at run time, this is considered to be
``reprogramming'' or ``retraining'', which is not considered as a run-time
tuning knob for energy-accuracy trade-off.

While operating both CNN and MLP, a large number of nodes is involved and as a
result the energy per classification operation is relatively high and can range
from tens of $nJ$ to several $\mu J$ per classification.\ignore{Similarly, in case of
SVMs the energy per classification increases with the kernel complexity,
which have similar energy per classification ranges. \fixme{might need to quantify}}
Similarly, energy dissipation of the SVMs range from several $nJ$ to tens of
$\mu J$ depending on the choice of the kernel function.
This high energy consumption per classification is tolerable in case the
energy budget is not limited (i.e.: desktop computer).
However, this is not acceptable in
energy-constrained scenarios, like when we need to run classification operation
on a mobile system. In such scenarios the classification operation generally
gets offloaded to the cloud or the classification is performed locally with
reduced accuracy goals.\ignore{This is also worsened by the fact that, CNNs, MLPs and
SVMs are trained for certain goal accuracy (usually maximum), and changing the
accuracy-energy-efficiency trade-off is challenging due to the inherent
complexity (for neural networks) or limited number of tuning knobs \missingcitation (for SVMs).} The inherent complexity of CNNs and MLPs, as well as the nature of the SVMs, makes it difficult to play the accuracy vs. energy-efficiency game at run-time. For example,
it is possible to turn off some of the nodes
in CNN and MLP, but it is not always obvious which of the nodes contribute the
most to the overall accuracy, and the behavior in that case might become
unpredictable (both in deep and shallow networks).}
\ignore{
\begin{figure}[tb]
  \centering
  \includegraphics[width=0.8\columnwidth]{rfvscnn.svg.pdf}
  \caption{Accuracy degradation for MLP and RF. For simplicity,
  a reduced \texttt{Digits} dataset was used. In the RF every
  estimator is a DT.}
  \label{fig:RFvsCNN}
\end{figure}
}

\ignore{Check if we need any info from the two paras that were deleted}RF is composed of binary decision trees ($DT$) (see Figure~\ref{fig:logical:rf-fog}a), and although the entire RF is composed of $O(t2^d)$
decisions, where $t$ is the number of trees, and $d$ is the upper bound on the tree depth, evaluation during testing requires only $O(td)$ computations.
Every $DT_i$
receives some input features $X_{Ri}$, where $X_{Ri}$ is a random subset of input $X$.
A ``Majority Vote'' across trees is then
used to identify the label. Such an approach avoids overfitting
and ensures high accuracy \cite{breiman}.\ignore{Algorithm \ref{alg:BudgetRF} shows an implementation
of budgeted RF as described by Nan et al. \cite{icml2015_nan15}. Algorithm described by Nan et al. \cite{icml2015_nan15} is an example of budgeted training.} Note that
during the random forest training, the trees are generated depending on their
validation cost, where the cost could be energy,
delay, energy-delay product, or accuracy.\ignore{The reason {\bf validation} cost is used as a metric is because we would like to make sure that the classification cost of ``unseen'' examples (and not ``training'') is minimized.}\ignore{
Because the DTs within RF work independently on random subsets of
input features, those decision trees can be seen as having additive properties,
which means that even if several of the DT blocks are turned off,
the total accuracy degrades gracefully (Figure \ref{fig:RFvsCNN}).
This ``scalability'' property, where
the accuracy of RF scales with the number of DTs makes
the RF a suitable candidate for budget-limited environments, with budget being
energy, delay, accuracy, or a combination of them.
For example, depending on the budget requirements, some of
the trees could be turned off to improve energy efficiency, but at the cost of
statistical performance. This type of ``scalability'' is also observed in CNNs
and MLPs but the scalability is less forgiving and hence, those algorithms are
hard to adapt to environments with changing budgets.
}\ignore{RF is composed of exchangeable DT's that in the limiting case, given enough
data, each represent a noisy estimate of the label for a new test example. By
combining outputs of multiple DT's in an RF, we reduce the variance of our
estimate.} Turning off DT blocks generally leads to a graceful degradation of
accuracy\ignore{(see Figure \ref{fig:RFvsCNN})}, as the predicted label for a new test example is independent, in contrast to CNN and
MLP, where each node in the network
is connected to many other nodes, and it is usually difficult to predict how each node affects the accuracy of the the neural network at run-time (it is possible to achieve that offline however). In general, when operating at unlimited energy
budgets, CNNs and MLPs generally provide higher accuracy than RF, but RF provides us an opportunity to game accuracy for energy-efficiency. In
Section~\ref{sec:evaluation}, we show more detailed comparison of the
classifiers.
\ignore{
\begin{algorithm}
\begin{small}
  \caption{Constructing Feature Budgeted Random Forest. \textbf{Validation} cost is the energy dissipation cost of classifying ``unseen'' examples.}\label{alg:BudgetRF}
  \begin{algorithmic}[1]
    \Require Training feature set $X$; Training label set $y$; Budget $B$ -
    budget could be accuracy, energy, delay, energy-delay-product, etc.
    \Ensure Set of trained decision trees $T$
    \Procedure{BudgetRF}{$X, y, B$}
      \State $T \leftarrow \emptyset$
      \Comment Assume that empty set has a cost of 0
      \While{Average {\bf validation} cost on $T \le B$}
        \State $T \leftarrow T \cup$\Call{Tree}{$X, y$}
      \EndWhile
      \State \Return $T$
    \EndProcedure
  \end{algorithmic}
  \end{small}
\end{algorithm}
}

%% file: groves.tex
\boxaround{
\begin{enumerate}
  \item {\bf Content}
  \begin{enumerate}
    \item Detailed description of the microarchitecture
    \item Detailed description of the construction algorithm -- Note that
      there is no separate training algorithm for the multi-groves.
      Training is inherited from the RFs
    \item Description of the testing phase and algorithm
    \item Description of the individual blocks of the groves
    \item How is the minimum size of a grove identified?
  \end{enumerate}
  \item {\bf Images}
  \begin{enumerate}
    \item Block diagram of the microarchitecture
    \item Block diagram of an individual grove
  \end{enumerate}
  \item {\bf Algorithms}
  \begin{enumerate}
    \item Groves construction algorithm \ref{alg:GCTrain}
    \item Testing algorithm \ref{alg:GCEval}
  \end{enumerate}
\end{enumerate}
}

\begin{algorithm}[t]
\begin{small}
  \caption{Constructing Field of Groves Classifier}\label{alg:GCTrain}
  \begin{algorithmic}[1]
    \Require Number of estimators $n > 0$; Maximum size of a grove $k \le n$;
    Training set $X, y$
    \Procedure{GCTrain}{$n,k,X,y$}
      \State Train $RF \leftarrow$ \Call{RandomForestTrain}{$n,X,y$}
      \State \Return \Call{Split}{RF, k}
    \EndProcedure
    \Subroutine{Splitting a Random Forest}
    \Require Pretrained random forest $RF$; Maximum size of a grove $k > 0$
    \Ensure Split grove ensemble $GC$
    \Procedure{Split}{$RF,k$}
      \State $i \leftarrow 0$
      \State $GC \leftarrow \emptyset$
      \While {$i < $ \Call{Length}{$RF.estimators$}}
        \State $G \leftarrow$ new Random Forest of size $k$
        \State $G.estimators \leftarrow RF.estimators[i..i+k]$
        \State $GC \leftarrow GC \cup G$
        \State $i \leftarrow i + k$
      \EndWhile
      \State \Return GC
    \EndProcedure
  \end{algorithmic}
  \end{small}
\end{algorithm}
\begin{algorithm}[t]
\begin{small}
  \caption{Evaluating Field of Groves Classifier}\label{alg:GCEval}
  \begin{algorithmic}[1]
    \Require Stopping threshold $0 < thresh < 1$; Maximum number of hops $max\_hops \le$ Number of groves $n\_groves$; Input set $X$
    \Procedure{GCEval}{$X, thresh, max\_hops$}
      \ParFor {every $x$ in $X$}
        \State $start\leftarrow$\Call{Random}{{\bf from} 0 {\bf to} $n\_groves$}
        \Comment Start at random $grove$ to avoid bias
        \State \texttt{prob} $\leftarrow \{0\}^{\#labels}$
        \For{$j\leftarrow$ {{\bf from} 0 {\bf to} $max\_hops$}}
          \State $index \leftarrow (start + j) \mod n\_groves$
          \State $\texttt{prob} \leftarrow \texttt{prob} + Grove(index).predict\_prob(x)$ \label{alg:GCEval:eval:begin}
          \State $\texttt{prob\_norm} \leftarrow \texttt{prob} / (j+1)$
          \If{\Call{MaxDiff}{$\texttt{prob\_norm}$} $\ge thresh$}
            \State \Return \texttt{prob\_norm}
          \EndIf \label{alg:GCEval:eval:end}
        \EndFor
        \State \Return \texttt{prob\_norm}
      \EndParFor
    \EndProcedure
    \Subroutine{Minimum Difference of Maximum Values}
    \Require Array $ar$
    \Procedure{MaxDiff}{$ar$}
      \State max1, max2 $\leftarrow$ \Call{TwoMaximumValues}{$ar$}
        \State \Return \Call{abs}{max1 - max2}
      \Comment In case of ``Multi-output classification'', \Call{min}{$\cdot$} function is called prior to returning
    \EndProcedure
  \end{algorithmic}
  \end{small}
\end{algorithm}

\subsection{Random Forests: FoG Implementation}
\label{sec:groves}
\subsubsection{Training and Evaluation Algorithm}
\label{sec:groves:algo}
As described in section \ref{sec:random-forests}, RF is suitable for
environments where accuracy could be traded-off for energy efficiency. The main
advantage of the RF approach stems from the fact that the accuracy of
RF\ignore{scales with the number of DTs in the RF} tends to improve with an
increase in the number of DTs. Moreover, DTs have few active computational nodes
during prediction, and the nodes are generally of very low computational
complexity. \ignore{Moreover, the DTs have lower number of active computational
nodes which also are generally of much lower complexity than those in CNN -
individual computing node in a DT is a comparator, while in neural networks the
most basic computing node is a multiply-accumulate (MAC) block.}
One of the disadvantages that RF experiences is ``over-utilization'' of the computational resources. Previous works have shown that large portion of input samples within datasets are far enough from the decision boundaries, and do not require complex classifiers \cite{takhirov:islped:2016,venkataramani:2015:sce}. Conventional RFs, however, lack the ability to allocate less computational resources for the inputs if desired. This problem could be solved by using only a limited number of trees, depending on the current confidence level.

In this section we propose a novel RF implementation called Field of Groves
(FoG), which avoids any unnecessary expending of energy on inputs with low
uncertainty. Each grove is composed of a random, non-overlapping subset of the
trees from the ``original'' RF.\ignore{Every grove is a smaller
version of the ``original'' forest with randomly sampled decision
trees.}\ignore{The FoG system could be viewed as a forest of forests ensemble.}
Figure~\ref{fig:logical:rf-fog}(b) shows the logical view of our proposed FoG
implementation of RF. The training of the FoG is described in Algorithm
\ref{alg:GCTrain} and is done offline. During this training phase,
a RF is first pre-trained (using Algorithm from~\cite{icml2015_nan15}),
and the
DTs are randomly split into groves. The splitting involves a simple
division of the forest into sets with $k$ DTs, where $k$ is the size of the
grove.

The label evaluation algorithm for the approach is shown in Algorithm
\ref{alg:GCEval}. Here, for every input $x \in X$, we compute the confidence score
using one of the randomly selected grove. Confidence in this context is defined
as the difference between the most probable and second most probable labels\footnote{In case the classification problem is ``multi-label'' or ``multi-output'', the \texttt{MaxDiff} returns the \texttt{Min} of the differences within the label. That means that confidence level is defined as the ``minimum difference of the maximum values''}
If the confidence is higher than the goal threshold, the computation for $x$ is
complete. Otherwise, $x$ and the current probability distribution is sent to the next grove, where the
probability array is recomputed again form $x$ and combined with the one received from the previous grove. That way
more groves contribute to the inputs with high uncertainty. This process is repeated
until either the threshold is exceeded or the the entire forest is evaluated.
Note the contrast between FoG and conventional RF evaluation: in FoG the
groves return probability distributions which are averaged out across groves; in
the conventional RF however the DTs return class predictions, which are later put to
a majority vote.

\subsubsection{Micro-architecture}
\label{sec:groves:microarchitecture}
The high-level architecture of the FoG implementation of RF is shown on Figure
\ref{fig:groves-block}. Here, the groves are connected in a circular fashion, with each grove being able to send its current inference to the next grove. To understand the operation of the system, let's consider an example with a
3-class (class A, class B and class C) problem and the threshold value set to $0.1$. Let us assume that the processor sends an input $X$ that has 5 features. When FoG receives this input, it is assigned an $id$, and is sent to one of the groves (say grove G0 in Figure~\ref{fig:groves-block}) through the accelerator Input Queue.

\begin{figure*}[htp]
  \centering
  \includegraphics[width=0.85\textwidth]{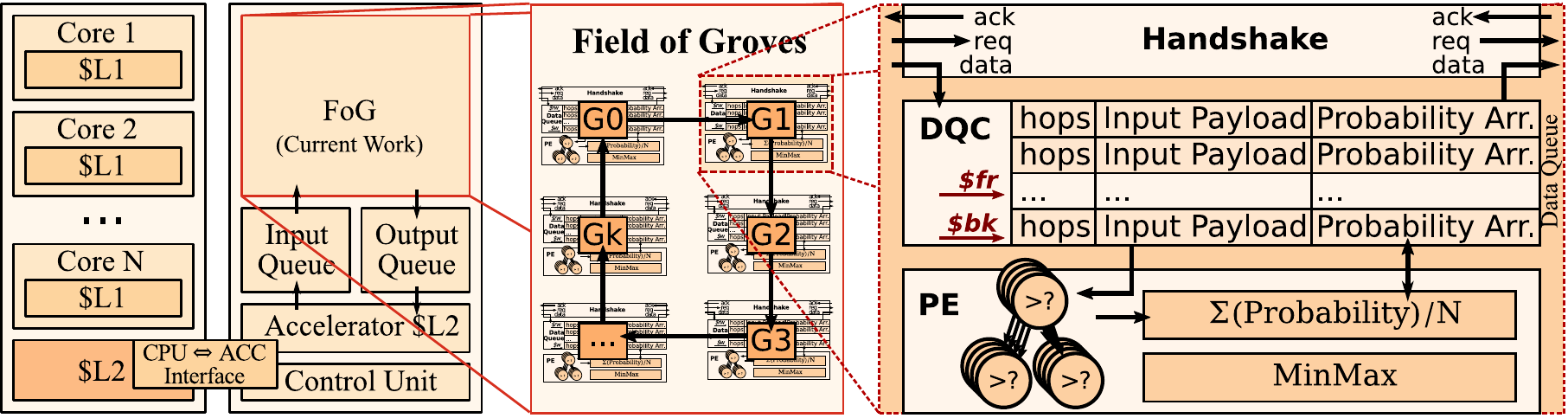}
  \caption{Random Forest implemented as Field of Groves and the microarchitecture
  of a grove. Notice the grove $G0$ can communicate to grove $G1$ through the ``Handshake'' block without going out of the FoG. The ``Data Queue'' includes a controller to maintain pointers $\$fr$ and $\$bk$. \ignore{Handshake controller makes sure the request and acknowledge
  signals are set to appropriate values when required. Local memory is distributed
  across all groves, and the Input Queue holds the current ``assignments'' for the
  grove. The memory fields structure is described in the later parts of Section
  \ref{sec:groves}}}
  \label{fig:groves-block}
\end{figure*}

{\bf Data Queue}\\
Once $G0$ receives a new input, it places it into the local memory, which serves as a data queue. The queue is controlled using two pointers: $\$fr_{G0}$ and $\$bk_{G0}$ for front and back of the queue respectively. $\$fr_{G0}$ always points to location that contains the input that is currently being processed and $\$bk_{G0}$ points to the first empty location at the back of the queue. For each input we store \texttt{Input Payload}, which holds the received input features + id; \texttt{Probability Array}, which contains the current prediction probabilities; \texttt{hops} which is a count of groves that have so far processed the current Input Payload. Whenever a new input is received: if the input is received from the processor, it is placed at the back of the queue. The \texttt{Input Payload} and \texttt{Probability array} values are set based on the information sent by the processor and the \texttt{hop count} is set to 0.
If the input is received from the neighboring grove, it is placed at the front of the queue. The \texttt{Input Payload} and \texttt{Probability array} values are set based on the information sent by the neighboring grove and the \texttt{hop count} is incremented by 1. This ensures that the input that were partially computed have higher priority. In our example, because the input is from the processor, it is placed into the $\$bk_{G0}$ location of the queue. For the new input: \texttt{\{hops = 0, Input Payload = $X$, Probability = \{0,0,0\}\}}

Data queue is controlled by the queue controller (DQC) which is responsible for maintaining the $\$fr$ and $\$bk$ pointers. For each received input, DQC routes $\$fr$ to the processing element, the processing element reads the entries corresponding to $\$fr$ and once the computation is complete, it writes the results back to $\$fr$. $\$fr$ and $\$bk$ are incremented by $\Gamma$, which is the length of queue word and represents number of rows in physical memory that are required to store the hop count, \texttt{Input Payload} and \texttt{Probability Array}. $\Gamma$ is a variable, and it depends on the number of features and number of classes in dataset. In our example, $\Gamma = 1 + 5 + 1 + 3 = 10$ (1 byte for \texttt{hops}, 5 bytes for features in \texttt{Input Payload} + 1 byte \texttt{id}, and 3 bytes to store the current label prediction\footnotemark in the \texttt{Probability Array}). Our current implemetation of the FoG has a data queue of 6kB, and can store 8 \texttt{MNIST} examples per grove. Note that the memory can be easily increased to support datasets with larger feature counts and label counts.\
\footnotetext{The reason every label is stored in a separate byte is because we use byte addressable memory to support reconfigurability.}

{\bf Processing Element (PE)}\\
\ignore{This will be in the figure}
The PE in every grove is represented by a set of decision trees and its operation is described in algorithm \ref{alg:GCEval}. The \texttt{Input Payload} ($X$) is processed by all the trees within the grove to determine the probability distribution of the labels. This result is then averaged with \texttt{Probability Array} received from previous grove or just written back  in case of a new input and the current confidence level is computed (as the difference between the two largest values in the \texttt{Probability Array}). The latency of the PE depends on the number of trees per grove, the maximum depth of each tree and degree of parallelism.

Once PE finishes the computation, a decision is made if the current confidence level is adequate. If so, the DQC is notified that the classification of the current \texttt{Input Payload} is complete and the computed result needs to be sent back to the processor. However, if the confidence level is lower than a threshold $thresh$, a request is sent to the next grove for further processing. Here the entire entry (\texttt{Hop Count, Input Payload} and \texttt{Probability Array}) for the current input is copied to the next grove. \ignore{Threshold value $thresh$ is a run-time parameter that is used as a tuning knob to trade-off accuracy for energy-efficiency.}

Continuing with the previous example, let us say that after $G0$ completes processing the input $X$, it returns the probability distribution of $\{0.32,0.35,0.33\}$. This is used to compute the confidence. In this example the confidence is $0.35-0.33 = 0.02$. Because the threshold was set at $0.1$, the classification of input $X$ is considered incomplete. It is written back to the location $\$fr_{G0}$, and a \texttt{req} flag in the handshake is raised. At this point, the $fr_{G0}$ is incremented, and grove $G0$ is ready for the next input. The value stored at $\$fr_{G0}-1$ is \texttt{\{hops = 1, Input Payload = $X$, Probability = $\{0.32,0.35,0.33\}$\}}

{\bf Handshaking Protocol}\\
Groves use a simple handshaking protocol to talk with each other. After $G0$ computes the output probabilities, it checks its confidence and if the confidence is low it sets a \texttt{req} flag to signal the neighboring grove $G1$ to copy the current input as well as computed probabilities. Once the copy is complete, an acknowledgment flag \texttt{ack} is raised by $G1$ for one cycle to notify that the copy procedure is complete. At that time $G0$ pulls the \texttt{req} line down, completing the handshake.

Because $G1$ receives its input from another grove (in our case $G0$), it places it at $\$fr_{G1}$ of its queue. Assume that $G1$ computed the probability distribution as $\{0.28, 0.45,$ $0.27\}$. These values are averaged with the values computed by $G0$. The entries corresponding to the current input are now \{\texttt{hops} = 2, \texttt{Input Payload} = $X$, \texttt{Probability} = $\{0.3,$ $0.4,$ $0.3\}$\} and the predicted label is $\argmax$ ${(\texttt{Probability})} = 1$. At this point, the threshold value constraint is met $0.4-0.3 \ge 0.1$, which indicates that this input does not require any further processing, and should be sent to the accelerator output queue. Note that in the example discussed above, the value of \texttt{hops} was increasing with every new grove.

\ignore{
\begin{figure*}[t]
  \centering
  \includegraphics[width=0.85\textwidth]{../prebuild/block_placeholder.jpg}
  \caption{Microarchitecture of two neighboring groves.}
  \label{fig:groves-block}
\end{figure*}
}
{\bf Run-time Tunability}\\
In our proposed FoG implementation of RF the energy-efficiency and accuracy
could be easily tuned by changing the \texttt{probability threshold} and
\texttt{maximum hops} parameters. The \texttt{threshold} parameter
indirectly controls the number of groves that process the input.
The \texttt{maximum
hops} parameter places an upper limit on the number of groves that process the input (based on EDP or accuracy constraints).
A detailed evaluation of how the probability threshold parameter and the maximum hop count parameter affects the energy efficiency and accuracy of our FoG implementation is presented in Section~\ref{sec:evaluation}. 

{\bf Reprogrammability}\\
\ignore{Anticipating future discussion on programmability of the FoG,}
To support various trained RFs corresponding to various datasets, the DTs were implemented to be reprogrammable. For a given dataset, every node is populated with the weights $\omega_i$, as well as memory address offsets for the respective features $x_j$. In addition to that the DQC is programmable to support variable step for the queue pointer.
For example, if a node $N$ checks the conditional $x_N > \omega_N$, then this node will store the constant $\omega_N$, as well as offset $OFF{x_N}$. This indicates that the location of the input $x_N$ is at $\$fr+OFF{x_N}$. At the same time the DQC stores a value $\Gamma$ and the next entry in the queue has an address $\$fr_{next} = \$fr + \Gamma$.
The reasoning behind having a variable step size $\Gamma$ is that we want to support different number of features as well as different number of labels for different datasets. For example, MNIST dataset has 784 features and 10 labels, while Penbase Digits dataset has only 16 features and 10 labels\footnote{Physically each entry of the data queue is spread over several rows. Here $OFF{x_N}$ is the offset within an Input}.

\ignore{Although our approach requires mechanism for queue management and needs an additional mechanism to ``dereference'' the offset pointer (which slightly increases the complexity per tree), this design choice is a necessary trade-off to support programming of hardware required to support datasets having different feature counts. In the current work, the other components of reprogrammable logic, such as configuration matrix upload and storage, dynamic routing optimization, and changes in the training routine were not considered, and will be addressed in the future work. \ajay{The last sentence will open up a whole bunch of questions.} \fixme{Do we need the last paragraph?}}

%% file: evaluation.tex
\section{Evaluation}
\label{sec:evaluation}
\input{experimental-setup}
\input{experimental-results}

%% file: experimental-setup.tex
\subsection{Experimental Setup}
\label{sec:experimental-setup}
\boxaround{
\begin{itemize}
  \item {\bf Content}
  \begin{itemize}
    \item Describe the setup and how the simulations are run
    \item Describe the flow of design
  \end{itemize}
  \item {\bf Images}
  \begin{itemize}
    \item Flow diagram similar to: {\url{http://vlsiarch.eecs.harvard.edu/wp-content/uploads/2016/05/reagen_isca16.pdf}}
  \end{itemize}
  \item {\bf Tables}
  \begin{itemize}
    \item Application datasets
  \end{itemize}
\end{itemize}
}

\ignore{
\begin{figure*}[t]
  \includegraphics[width=\textwidth]{design-flow.svg.pdf}
  \caption{System design flow diagram. At steps \circled{1} and
  \circled{2} basic building blocks are designed and simulated to acquire the
  Power-Performance-Area (PPA) numbers. The PPA is then fed into the step
  \circled{3} to get the accelerator configuration given energy constraints.
  The PPA numbers are also fed into the $\mu$Architecture exploration \circled{4},
  which generates a suitable chip layout for simulation \circled{5}}
  \label{fig:design-flow}
\end{figure*}
}

We designed SVM with linear regression kernel (SVM$_{LR}$), SVM with
Radial-Basis Function kernel (SVM$_{RBF}$), MLP, CNN, RF and FoG classifiers for
our analysis. To compare the different classifiers we used five different
datasets from the UCI library \cite{uci}, and the list of these datasets is
shown in table \ref{table:datasets} under the ``Dataset'' column. These datasets
were chosen because they represent a diverse set of workloads typically seen on a
mobile device.\ignore{During the training phase, 60\% of the inputs were used for
training, 30\% used for cross-validation, and the remaining 10\% for evaluation.} All the results shown in this section are acquired using the\ignore{
evaluation portion of the inputs, thus guaranteeing that} inputs never seen
before by the systems under test.

We used the following design flow for performing a detailed power, performance
and area comparison of our proposed FoG with other ML classifier algorithms:

  {\bf Step 1:}\ignore{Building a library of basic computational units.} First, basic computational blocks, such as adders, multipliers,
  multiply-accumulate (MAC), sigmoid, etc. that are required by all the
  classifiers are designed considering trade-offs between energy and delay by
  sweeping through architectural and circuit level parameters, such as
  bitwidth precision, parallelization, pipelining, etc. We used Aladdin tool
  \cite{aladdin} to explore the architectural design space, and Cadence tools to\ignore{design the building blocks. We also} extract Power-Performance-Area
  (PPA) values for each block in this step.

  {\bf Step 2:} Once the library of computational units is generated, it is
  used in the offline budgeted training described in~\cite{icml2015_nan15} and algorithm \ref{alg:GCTrain}.\ignore{Budgeted training is a modified
  training routine that iteratively converges to maximum possible accuracy while
  maintaining the validation cost within energy budget constrains. For example, if
  energy budget is set to be $B$ Joules per classification, and if during the
  validation the average energy per classification was estimated as $E$ for
  the current design, then the current design will be considered if and only
  if $E \le B$.} We used energy-delay product (EDP) as budget metric during this phase. If there are several designs that meet the energy constraints, we choose the one with the maximum accuracy.
  Budgeted training requires
  information about the costs of building blocks which is provided by the PPA
  models\footnotemark. We use SciKit-Learn \cite{scikit-learn}  for the
  training (and exploration of logical structure) of the classifiers.
  \footnotetext{Note that the cost could be defined as either energy, delay, area, accuracy or any combination of them. The PPA library has information about delay, energy, and area, while the accuracy cost is determined using
  cross-validation data.}
  
  {\bf Step 3:}\ignore{After the logical structure of the accelerator is determined,} At this step,
  the detailed hardware  microarchitecture of the accelerator is designed.
  Microarchitecture exploration is independent of training, and is done using
  Aladdin toolset \cite{aladdin}. The PPA models from the previous step are used
  during this step to determine Pareto optimal frontier and select the most
  energy-efficient design.

  {\bf Step 4:} In this final step we design the whole architecture using
  Chisel HDL \cite{chisel}. This design environment was chosen, as it
  generates both hardware description code (Verilog) as well as C++ functional
  model. That allows for the functionality of the hardware to be verified
  against software implementation for correctness. The Verilog code was
  synthesized using 40 $nm$ Global Foundries technology with Synopsys standard
  cells for detailed power-performance analysis.

\ignore{
After the training phase, the logical pre-trained structure was extracted from the SciKit-Learn. For
example, if after the training phase a suitable logical structure of a random forest was found by
SciKit-Learn, all the decision tree structures were extracted and implemented in the next phase, which
is the microarchitecture exploration. This step in the process is covered by the Aladdin tool which
sweeps through different architectural parameters in order to find a Pareto optimal solution for
the architecture. The goal at this step was to maximize the Energy-Efficiency which we define as
energy-delay product. Note that the training routine is not part of the microarchitecture because
the focus of the current work is on energy-efficient evaluation rather than training.
}

\begin{figure*}[t]
  \centering
  \subfloat[][ISOLET\label{fig:dt-sweep:isolet}]{\includegraphics[width=.2\textwidth]{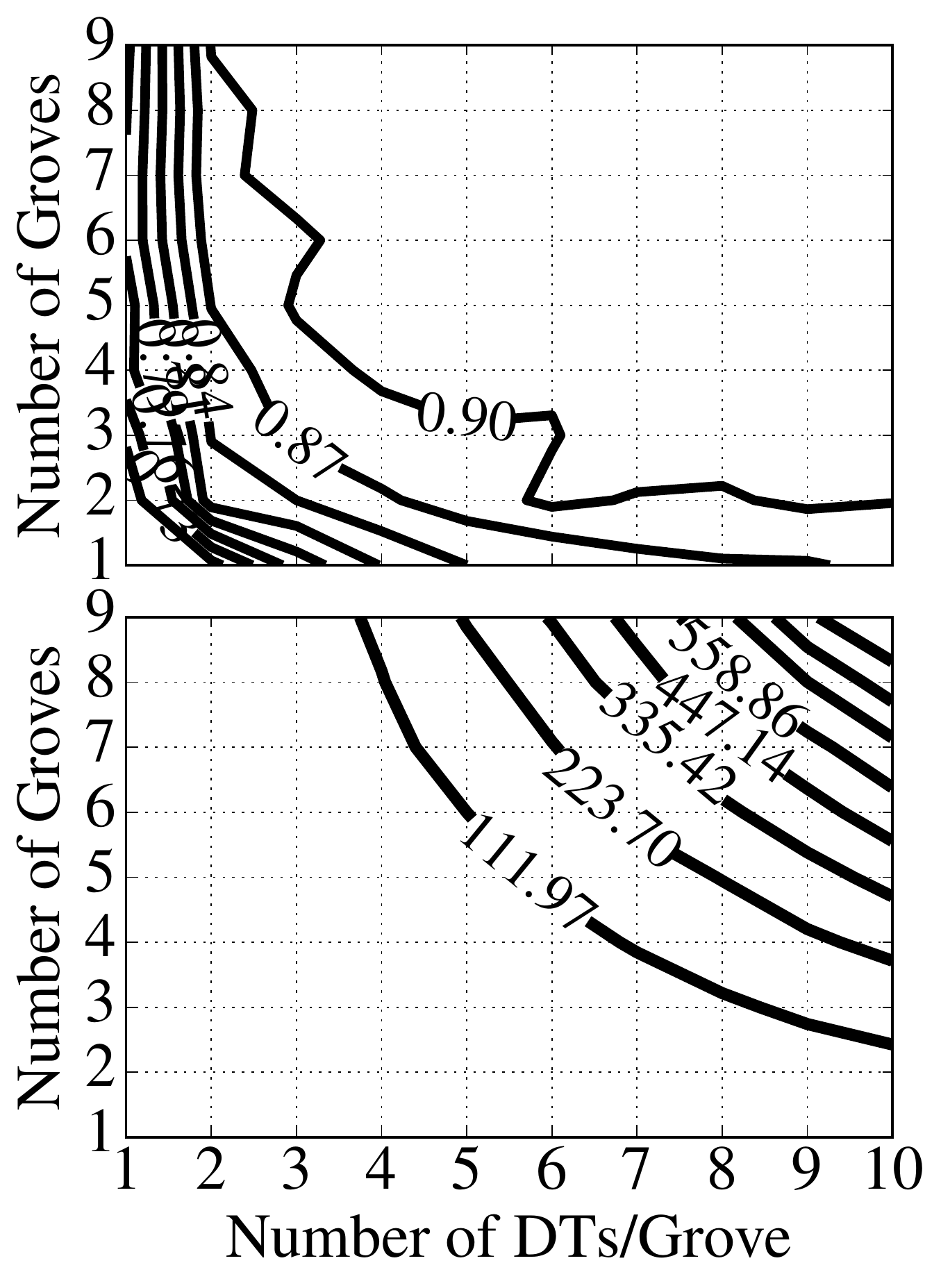}}
  \subfloat[][Penbase\label{fig:dt-sweep:penbase}]{\includegraphics[width=.2\textwidth]{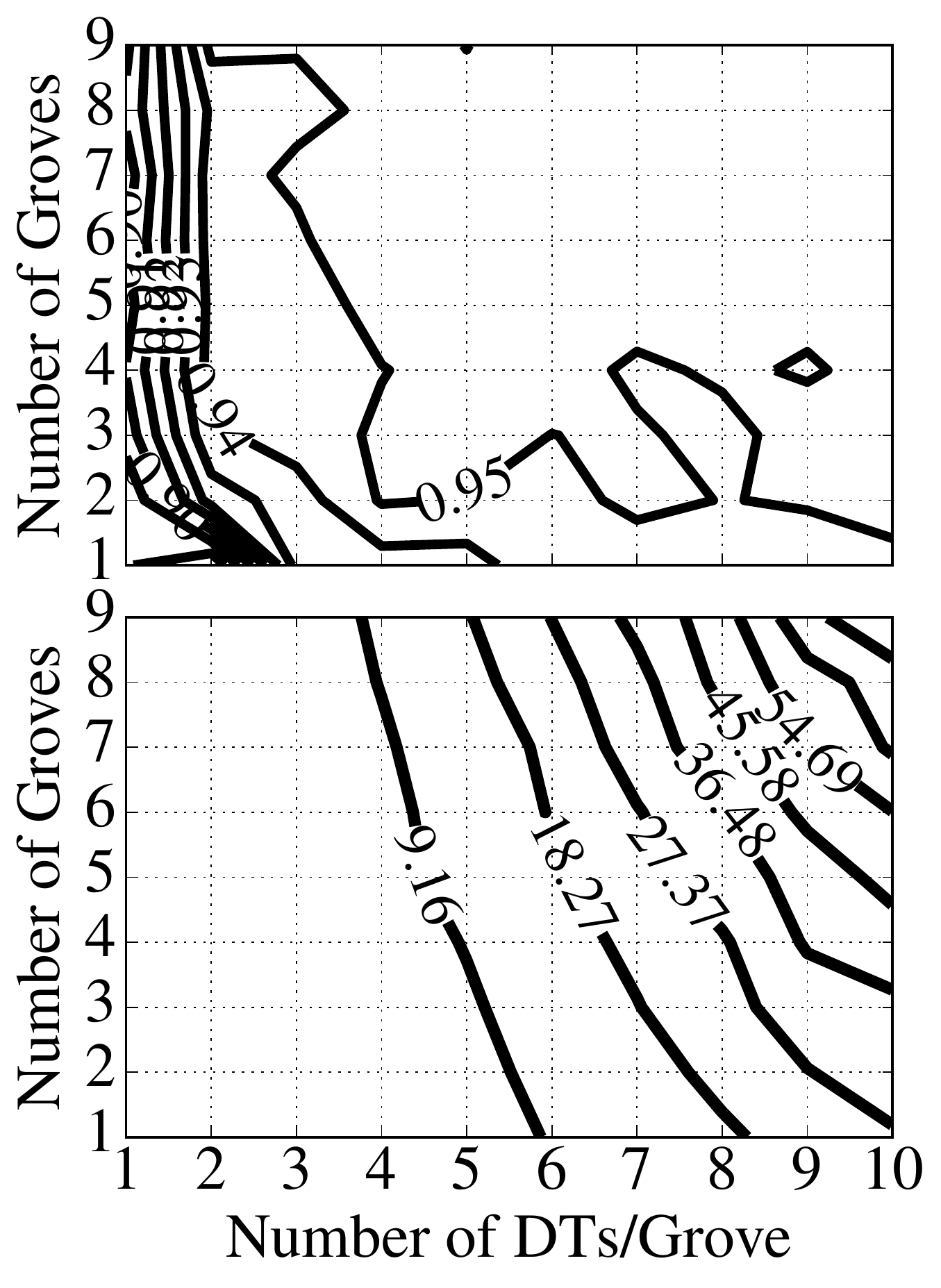}}
  \subfloat[][MNIST\label{fig:dt-sweep:mnist}]{\includegraphics[width=.2\textwidth]{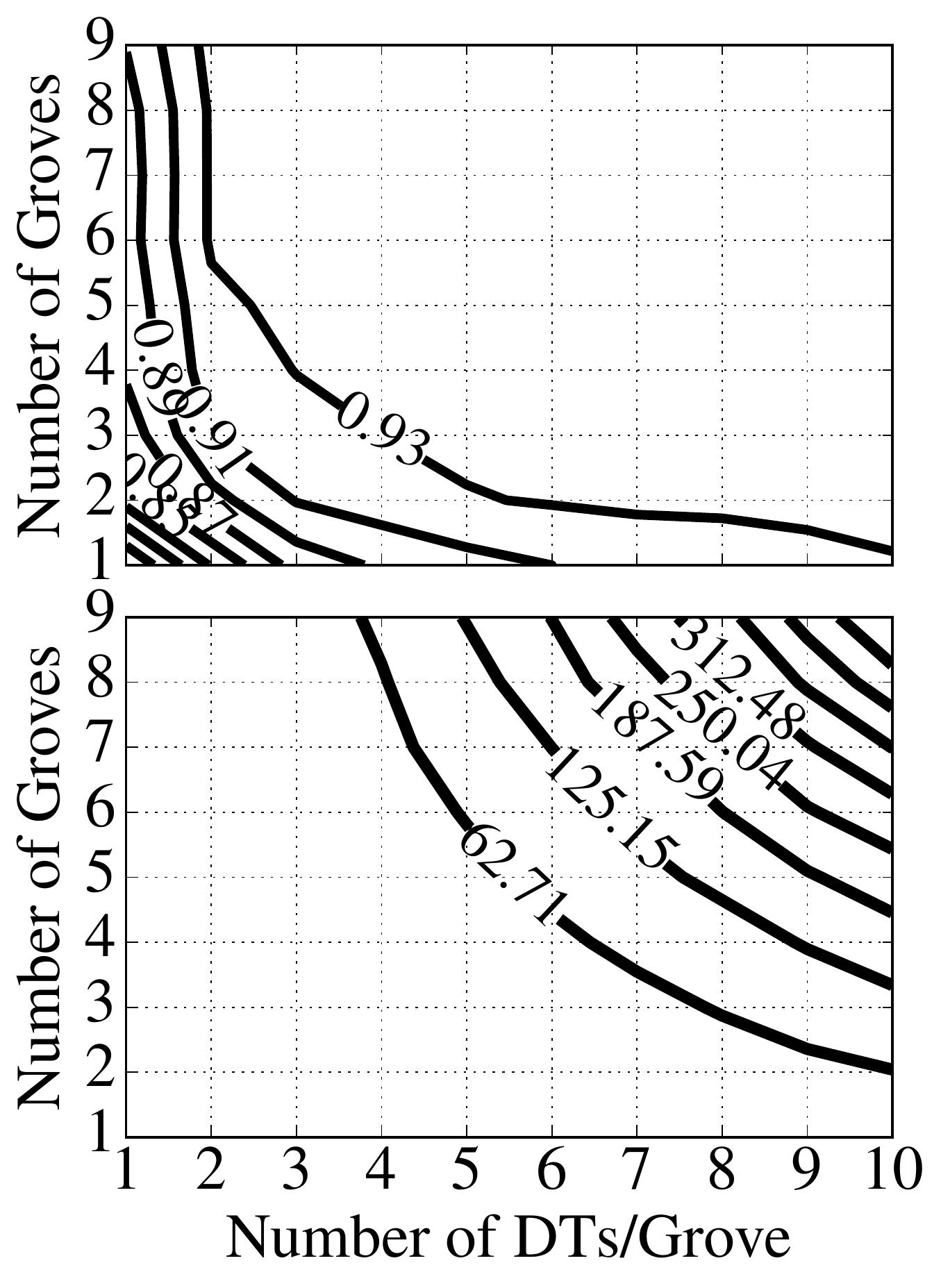}}
  \subfloat[][Letter\label{fig:dt-sweep:letter}]{\includegraphics[width=.2\textwidth]{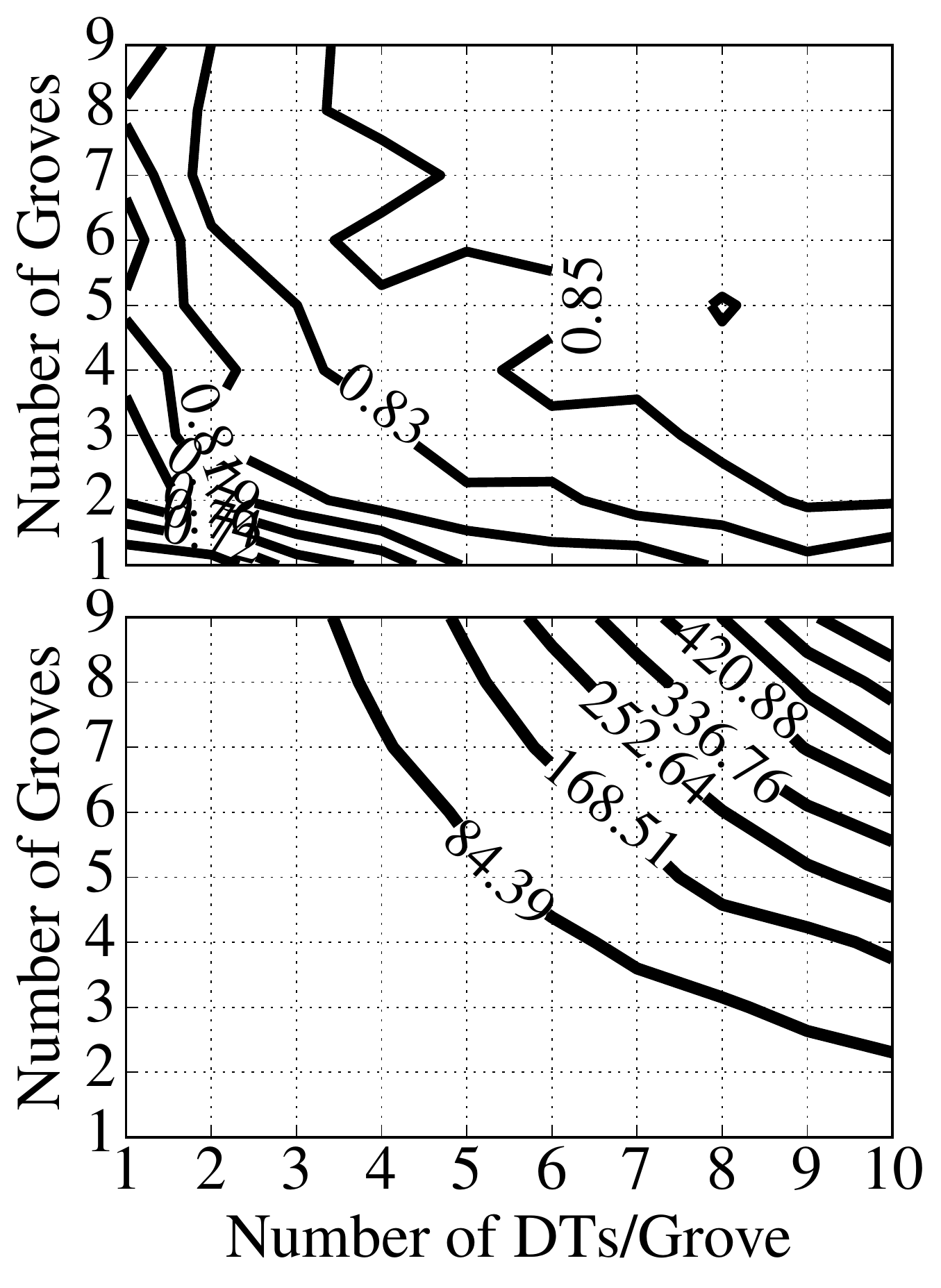}}
  \subfloat[][Segmentation\label{fig:dt-sweep:segmentation}]{\includegraphics[width=.2\textwidth]{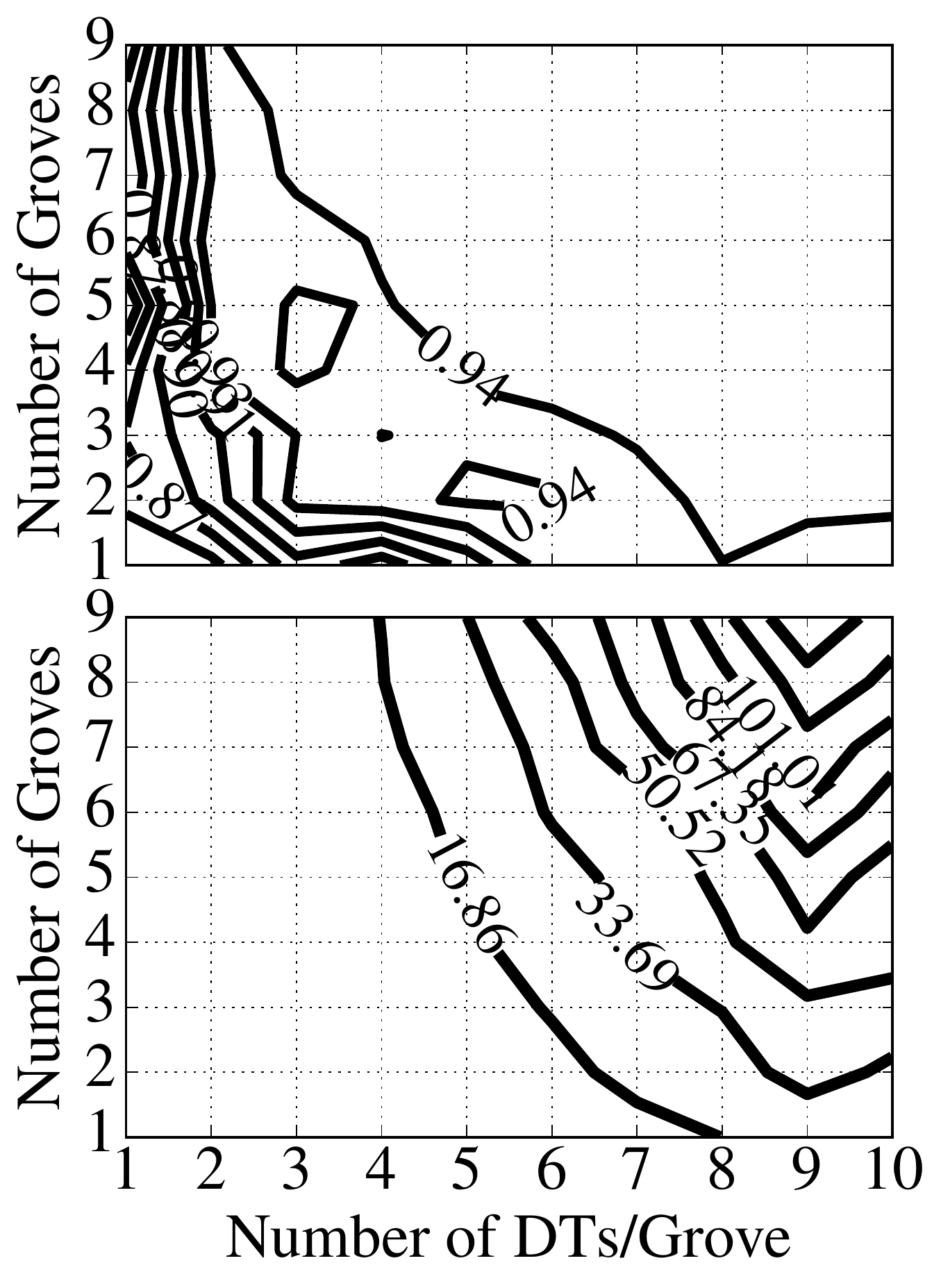}}
\caption{Accuracy and EDP as a function of ``Number of Groves'' and the ``Number of Decision Trees per Grove''. The product of the two variables shows the total number of decision trees in the FoG.}
  \label{fig:dt-sweep}
\end{figure*}

\textbf{FoG Design Considerations}\\
In our implementation, all the classifiers were designed for minimum EDP at maximum accuracy. The FoG classifier
was designed from the RF classifier by extracting the pre-trained DTs and re-assembling them into groves.
As described above, the number of decision trees per grove and the number of groves is decided during the design time. During the design time we analyzed the EDP and accuracy of different FoG topologies and the minimum EDP design point was selected (while maintaining the accuracy). Figure \ref{fig:dt-sweep} shows the accuracy and EDP results across different combinations of sizes of groves and total numbers of groves in the FoG.

\begin{figure}[t]
  \centering
  \subfloat[][8 Groves, 2 DTs/Grove\label{fig:tuning:8-2}]{\includegraphics[width=.24\textwidth]{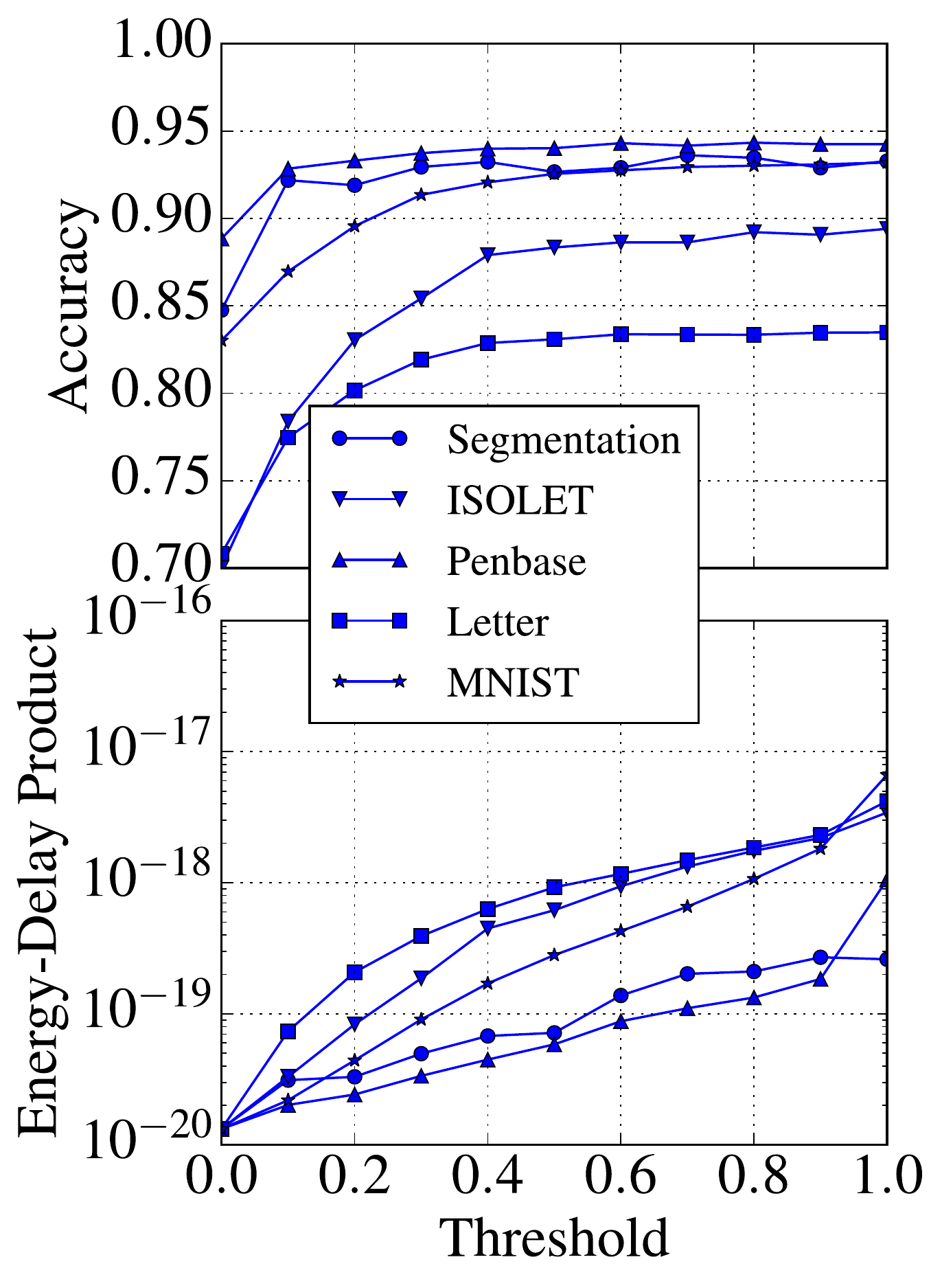}}
  \subfloat[][4 Groves, 4 DTs/Grove\label{fig:tuning:4-4}]{\includegraphics[width=.24\textwidth]{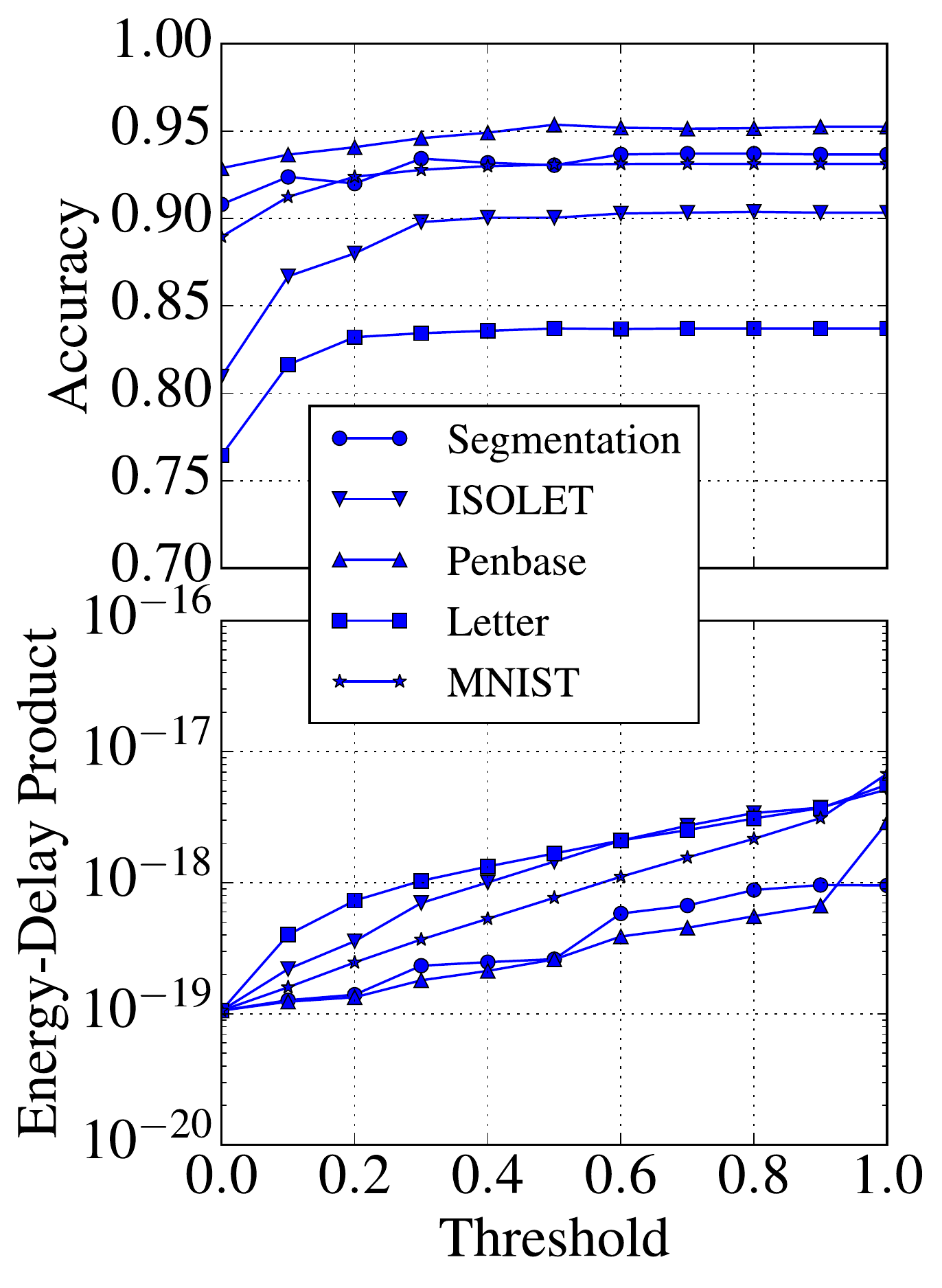}}
\caption{Example of FoG run-time tuning using the ``threshold'' variable.}
  \label{fig:tuning}
\end{figure}

To illustrate the choice of design time parameters while considering run-time tunability, let us discuss an example with 16 decision trees and \texttt{ISOLET} dataset. After examining the accuracy and EDP of different topologies (see Figure \ref{fig:dt-sweep:isolet}), we isolated two candidate topologies: 8x2 and 4x4\footnotemark. At this point we can use the ``run-time tunability'' as a deciding factor between these two roughly equivalent candidate topologies. Figure \ref{fig:tuning} shows the accuracy and EDP across all datasets as a function of threshold. Figure \ref{fig:tuning:8-2} shows that 8x2 topology is more energy-efficient, but the accuracy is lower for lower threshold settings. Figure \ref{fig:tuning:4-4} shows, in contrast, that the accuracy penalty for lower thresholds is not as drastic, but the energy-efficiency penalty is higher. In our case we go with 8x2 topology as minimum EDP is our primary goal. Note that once the physical topology is selected, the ``threshold'' variable could be changed during run-time to achieve a different operating point.
\footnotetext{We use $a$ x $b$ to describe a FoG topology with $a$ number of groves with $b$ decision trees in each grove.}

\ignore{Note that design-time
parameters differ for every evaluated dataset and the microarchitectures were
optimized per dataset.}

\ignore{
\subsection{Functional Modeling}
\label{sec:functional-modeling}

\boxaround{
This section will describe the functional modeling

Instead of simulating the sweep to find the architecture with goal parameters:
\begin{enumerate}
  \item Get the energy/delay numbers for the smallest building blocks through simulations
  \item Identify the settings for the algorithms using SciKit
  \item Estimate the energy/delay by plugging the numbers from building blocks into the architecture
\end{enumerate}

The preliminary results for basic building blocks:
\begin{itemize}
  \item {\bf Floating point comparison}
  \begin{itemize}
    \item {\em Energy:} 3e-12
    \item {\em Delay:} 370e-12
  \end{itemize}
  \item {\bf Floating point addition}
  \begin{itemize}
    \item {\em Energy:} 2.3e-12
    \item {\em Delay:} 290e-12
  \end{itemize}
  \item {\bf Sigmoid (LUT)}
  \begin{itemize}
    \item {\em Energy:} 6.8e-12
    \item {\em Delay:} 450e-12
  \end{itemize}
\end{itemize}

What do we need in this section?

\begin{itemize}
  \item {\bf Content}
  \begin{itemize}
    \item What is the need for modeling?
    \item How can we model the energy/delay/accuracy with maximum reliability?
  \end{itemize}
  \item {\bf Figures}
  \begin{itemize}
    \item Figure \ref{fig:modeling_lsqe} and similar to show the MSqE of the modeling approach
    \item Figure similar to the one in {\url{http://vlsiarch.eecs.harvard.edu/wp-content/uploads/2016/05/reagen_isca16.pdf}}, where they show different ML algorithms and accuracy
    \item Contour plot for Neural Nets - width vs. height vs. accuracy {\bf Maybe in later section}
    \item Contour plot for RF/Groves - size vs. depth vs. accuracy vs. EDP {\bf Maybe in later section}
  \end{itemize}
  \item {\bf Tables}
  \begin{itemize}
    \item Table showing the discrepancy of modeling and simulation
  \end{itemize}
\end{itemize}
}

\begin{figure}[t]
  \includegraphics[width=\columnwidth]{model_rf}
  \caption{Modeling vs. Simulation results for \texttt{Image Segmentation} dataset. Modeling is done using 3rd order polynomial LSQE for $x = [1,4,8,12]$}
  \label{fig:modeling_lsqe}
\end{figure}
}

%% file: experimental-results.tex
\subsection{Experimental Results}
\label{sec:experimental-results}
\boxaround{
\begin{itemize}
    \item {\bf Algorithms to consider}
    \begin{itemize}
        \item RF/Groves
        \item CNN/MLP
        \item SVM
        \item KNN
    \end{itemize}
    \item {\bf Platforms to run. Note: All architectures HAVE TO be finished a week before deadline to give enough time for simulations}
    \begin{itemize}
        \item Software on multithreaded machine (SciKit)
        \item Custom implementation in Chisel (+ functional verification vs. C++)
        \item GPU simulation
        \item FPGA simulation (RISC-V)
    \end{itemize}
    \item {\bf Simulations to run}
    \begin{itemize}
        \item Iso-energy: Keep energy ~fixed, sweep accuracy
        \item Iso-accuracy: Keep accuracy ~fixed, sweep EDP
    \end{itemize}
    \item {\bf Tables}
    \begin{itemize}
        \item Accuracy results cross-compared for maximum accuracy
        \item Energy results cross-compared for minimum energy
        \item Utilization for different datasets
    \end{itemize}
    \item {\bf Figures}
    \begin{itemize}
        \item Energy-efficiency vs. Accuracy - compares the tunability of the system
        \item Dual plots Utilization vs. \{Accuracy, EDP\}
        \item Layout screenshot?
    \end{itemize}
\end{itemize}
}

\ignore{
\begin{table}[tbh]
    \begin{tabular}{lccc|cc}
        \toprule
        ~           & MLP       & CNN       &   RF      & \multicolumn{2}{c}{Groves}    \\
                    \cmidrule(l){2-4}                   \cmidrule(r){5-6}
                    &                 & Accuracy  &           & Accuracy  & Utilization   \\
        \midrule
                      Image Segm.     & 90\%      & 93\%      & 92\%      & 90\%          & .7 \\
                      Spam            & 88\%      & 98\%      & 96\%      & 95\%          & .6 \\
                      Iris            & 95\%      & -         & 98\%      & 97\%          & .6 \\
                      Pen Base        & 83\%      & 97\%      & 94\%      & 88\%          & .8 \\
        \bottomrule
    \end{tabular}
    \caption{(Software - SciKit) parameters: threshold = 0.6, estimators = 12, max grove size = 2, max depth = 5; MLP: Threshold function, 6+6 nodes, CNN: Sigmoid, 2x2 Square kernel}
\end{table}
}


To perform a comparison of the classifier algorithms listed in section \ref{sec:experimental-setup}, we first trained all the algorithms for their maximum accuracy
without worrying about energy efficiency. Table \ref{table:datasets} shows the
comparison of accuracy between different classifiers. Two different numbers for the FoG are reported: FoG$_{max}$ and FoG$_{opt}$. FoG$_{max}$ shows the results for the FoG with its ``threshold'' parameter set to maximum. This forces the FoG to behave like an RF because every input will have to go through every decision tree of every grove. FoG$_{opt}$ shows the results for the case when confidence threshold was set to accuracy optimal point -- a threshold point above which accuracy does not increase with threshold but below which accuracy decreases with decrease in threshold.

From the table we can see that CNN has the highest accuracy for all datasets. The accuracy of the
traditional RF classifier is comparable to CNN for all datasets.
In terms of energy per classification, RF consumes $\approx$15$\times$, $\approx$1.7$\times$, and
$\approx$23.5$\times$ less energy than SVM$_{RBF}$, MLP and CNN, respectively. The RF energy
consumption is $\approx$10$\times$ higher than that of SVM$_{LR}$, but RF on average provides 20\%
higher accuracy than linear SVM. The very low energy dissipation in RF is due to
the fact that the basic computational unit in a DT is very simple (a basic comparator).

Table \ref{table:datasets} also shows the accuracy and energy dissipation of our
proposed FoG implementation of the RF classifier. Here all classifiers have been designed to operate at 1 $GHz$. The maximum achievable
accuracy of the FoG (both $max$ and $opt$) implementation is lower than RF and CNN by 3.2\% and 4\%,
respectively, but FoG$_{opt}$ classifier consumes $\approx$1.5$\times$ and $\approx$34.7$\times$ lower energy than RF and CNN, respectively. The FoG$_{max}$ on average consumes almost the same amount of energy energy as RF, and $\approx$23$\times$ lower than CNN.
When comparing to the SVMs, the accuracy of the FoG classifier outperforms the linear support vector machine SVM$_{LR}$ by $\approx$15\% on average, and achieves comparable statistical performance when compared to the SVM$_{RBF}$. In terms of energy SVM$_{LR}$ is $\approx$10$\times$ more efficient on average, while SVM$_{RBF}$ is more expensive ($\approx$23.6$\times$ higher energy consumption when compared to FoG$_{opt}$).

The main advantage of the FoG is that while achieving statistical performance comparable to the performance of the RF, it also allows easy run-time change in the energy-accuracy trade-off. Figure \ref{fig:tuning} shows how accuracy could be traded off for energy for 8x2 and 4x4 FoG topologies. Figure \ref{fig:tuning:8-2} shows that energy-efficiency could be easily improved by an order of magnitude without sacrificing much accuracy by tuning the confidence threshold from $1.0$ to $0.5$ for most datasets. After that a ``trade-off'' region of tunability starts -- one can improve energy-efficiency by trading off accuracy. This run-time tuning opportunity will prove beneficial in environments with constraint energy. The figure shows that for 8x2 design, two orders of magnitude improvement in energy efficiency could be achieved by tuning the confidence threshold from $0.5$ to $<0.1$. The accuracy drop in case of aggressive confidence tuning is anywhere between 10\% to 30\% depending on the dataset. The story is similar for 4x4 topology (figure \ref{fig:tuning:8-2}), however, the ``trade-off'' region of tunability starts at confidence threshold of $\approx$0.3. Although the accuracy drop is not as drastic, the EDP for 4x4 topology is much higher -- an order of magnitude higher for low accuracy, and equivalent for high accuracy points.

\ignore{
\begin{figure}
    \centering
    \subfloat[][Run-time\label{fig:tuning:run-time}]{\includegraphics[width=.48\columnwidth]{acc_edb_thresh.svg.pdf}}
    \subfloat[][Design-time (ISOLET)\label{fig:tuning:design-time}]{\includegraphics[width=.48\columnwidth]{ISOLET-FoG_dt.svg.pdf}}
    \caption{Change in accuracy and EDP as a function of run-time and design-time variables. (a) shows that ``threshold'' could be used for dynamic tuning;\ignore{Notice that the accuracy ``saturates''.} (b)Accuracy and EDP as functions of design-time parameters: ``Number of groves per classifier'' and ``Number of trees per grove'' for ISOLET dataset}
    \label{fig:tuning}
\end{figure}
}
\ignore{
it is possible for the end user to trade off energy
for accuracy during runtime. The run-time tunability of the FoG implementation
is shown on the contour plots in Figure \ref{fig:contour-grove}. The x-axis on
these plots represents the design-time variable (total number of decision
trees), while y-axis is a run-time variable (grove confidence threshold). At run
time, by controlling the\ignore{number of DTs that are used for
classification} confidence threshold, we can easily trade-off accuracy for energy efficiency. As an
example, suppose that the design of the FoG has 14 DTs, with 2 DTs per grove. At the run
time suppose the mobile system could be connected to a reliable power supply
i.e. there was no limit on the energy budget, we could set the confidence
threshold to the levels of maximum accuracy. Looking at Figure
\ref{fig:cont-mnist}, with 14 DTs we can set the threshold anywhere
between $0.8$ to $1.0$ to achieve maximum accuracy. The energy dissipation in
that case would be $1.5 nJ$ to $6.0 nJ$ per classification. In the case where
the mobile system was running of its battery, i.e. there is a limited energy
budget, and the user can control the grove confidence threshold to tradeoff
accuracy for energy efficiency For example, if the energy budget is $<1.5$ $nJ$
per classification, the confidence threshold is dropped to $<0.6$, which would
cause the accuracy to drop to $\approx 80\%$. Such precise and simple tuning is
not really possible in other classifiers. For instance, neural-networks can
trade-off accuracy for energy-efficiency by physically turning off some of the
neurons, however, that requires a mechanism to estimate which neurons would be
most beneficial to turn off. And although it is possible to do that offline,
this mechanism can be expensive and the granularity of such control can be very coarse
\missingcitation\fixme{Need reference!!!}.
}

\ignore{Need a paragraph discussion on how does accuracy/energy change with the number of groves and DTs!!! This is the same as the Throughput discussion!!! Reference figure \ref{fig:throghput}}
\ignore{
\begin{figure}[htp]
    \centering
    \includegraphics[width=0.95\columnwidth]{groves_block.svg.pdf}
    \caption{\fixme{This is a placeholder for discussion on design-time variables sweep and tradeoffs. Ignore it for now1!}}
    \label{fig:throughput}
\end{figure}
}
\ignore{
\begin{figure*}[htp]
    \centering
    \includegraphics[width=0.99\textwidth]{fog_new_new.svg.pdf}
    \caption{Accuracy vs. Energy-Delay Product (logscale) $\times 10^{-19}$ for RF, CNN, MLP and SVM$_{RBF}$. The values are acquired by sweeping through their design time variables -- number of estimators within an RF, and number of layers and number of nodes per layer for neural networks. SVM$_RBF$ was not sweeped, and is shown for comparison.}
    \label{fig:plots-comparison}
\end{figure*}
}



\ignore{
\begin{figure}[htp]
    \centering
    \subfloat[\texttt{ISOLET}]{
        \label{fig:cont-isolet}
        \includegraphics[width=0.75\columnwidth]{ISOLET-latency.svg.pdf}}\\
    \subfloat[\texttt{Penbase}]{
        \label{fig:cont-penbase}
        \includegraphics[width=0.75\columnwidth]{Penbase-latency.svg.pdf}}\\
    \subfloat[\texttt{MNIST}]{
        \label{fig:cont-mnist}
        \includegraphics[width=0.75\columnwidth]{MNIST-latency.svg.pdf}}\\
    \subfloat[\texttt{Letter}]{
        \label{fig:cont-letter}
        \includegraphics[width=0.75\columnwidth]{Letter-latency.svg.pdf}}\\
    \subfloat[\texttt{Segmentation}]{
        \label{fig:cont-segmentation}
        \includegraphics[width=0.75\columnwidth]{Segmantation-latency.svg.pdf}} 
    \caption{FoG Accuracy and Energy per classification for different datasets. \ignore{The grove size is set to 2}The number of decision trees per grove is set to 2,
    while the {\bf x-axis} is the design-time variable (number of DTs in the system);
    {\bf y-axis} is the run-time variable (grove threshold). \ignore{Do you expect similar behavior if you had 3 or 4 groves?}}
    \label{fig:contour-grove}
\end{figure}
}
\ignore{
A comparison of FoG implementation of RF with the traditional implementation of
RF and CNN using accuracy and energy efficiency metrics is illustrated in Figure
\ref{fig:plots-comparison}. FoG (5) and FoG (9) are two example designs with 5
and 9 groves respectively, where every grove consists of two DTs.
\ignore{I don't understand what you have written in the brackets.} The reason
these design parameters were used is because through simulation sweep we have
determined that for the given datasets 9 groves of size 2 was sufficient to
achieve the maximum accuracy.

The plots for FoG (5) and FoG (9) were generated by
tuning the run-time parameter (confidence threshold). The other plots shown on
the same figure are for traditional RF and CNN. These plots were generated by
sweeping through design-time parameters such as the depth and the width of CNN
and the size and depth of the RF. From the plots we can see that RF outperforms
the CNN by 2 orders of magnitude in terms of EDP, while maintaining the same
accuracy levels. The FoG implementations are slightly more energy-efficient than
RF. Moreover, it provides the end user with the flexibility to tradeoff accuracy
for energy-efficiency at run time, which is not possible in traditional RF and
CNN.
}
Table \ref{table:datasets} also shows the area comparison between different
classifiers. It must be noted that most classifiers' area changes
drastically with the internal parameters -- e.g. convolutional layers sometimes
implemented as having ``volume'' activation\ignore{do we need to explain that -- this is regular in CNNs},
and changing the size of one layer,
might contribute to the total area change cubically. Overall, the area of our
FoG implementation is larger than all classifiers except CNN.

\ignore{
\ajay{Do we use the same threshold for all groves?}

\zafar{Yes, the threshold is a global controlled externally by whoever sets the energy budget -- maybe a processor}

\ajay{Also, can't we also control the number of groves
that we use at run time?}

\zafar{No, the number of groves or the size of the groves are design time parameters. We could do it
in the future, but the control system would become uber complex}

\ajay{How would the accuracy compare if we use 2 groves with
very high threshold versus 4 groves with 0.5 threshold? Or are you indirectly
controlling the number of groves used for classification by controlling the
threshold?}

\zafar{Please refer to figure \ref{fig:contour-grove}}
}

\ignore{
By utilizing this property of the FoG implementation of the RF we can sacrifice
accuracy to regain the energy efficiency. Simulation results show that depending on the
number of groves in a FoG, we can scale the energy efficiency linear to the number of
groves, while the accuracy drops non-linearly with the number of groves in a FoG.}

\ignore{
Looking back at table \ref{table:datasets}, one can see that for the given datasets the
Random Forest approach achieves order of magnitude lower energy dissipation as compared to
CNNs. The FoG variation of the RF implementation fails to achieve accuracy as high as RF or
CNN, however, looking at figure \ref{fig:plots-comparison} we can see that we can
trade-off energy for accuracy at runtime. The amount of tuning, as seen from the figure,
depends on datasets and physical implementation.

Comparing the plots on Figure \ref{fig:plots-comparison}, we note that the RF approach is
capable of achieving orders of magnitude lower energy-dissipation while maintaining the accuracy
levels.

 This is illustrated on figure \ref{fig:contour-grove}.
X-axis on the figure is a design-time parameter (number of DTs), while the y-axis is the
run-time parameter (threshold), that controls if more computation is required to achieve higher accuracy.
Figure \ref{fig:plots-comparison} also show two different implementations of FoG classifiers:
FoG (5) has 5 groves, while FoG (9) has nine groves. The different points on the figure were achieved
by tuning the runtime-parameter. These plots show the random forest classifier compared to the
(convolutional) neural network. Different accuracy/EDP numbers were acquired by sweeping through
design time variables like depth and width of the network, number of DTs in the random forest,
and maximum depth of each individual tree. All these parameters affect accuracy and energy dissipation.

\footnotetext{Note that the proposed work was not evaluated against deep networks because
the current work is focused more on computing under tight energy constraints.}

Looking back at table \ref{table:datasets}, one can see that for the given datasets the
Random Forest approach achieves order of magnitude lower energy dissipation as compared to
CNNs. The FoG variation of the RF implementation fails to achieve accuracy as high as RF or
CNN, however, looking at figure \ref{fig:plots-comparison} we can see that we can
trade-off energy for accuracy at runtime. The amount of tuning, as seen from the figure,
depends on datasets and physical implementation. For example, for the \texttt{Penbase} figure
FoG (5) we can see that FoG (5) is tunable during runtime, and it is possible to reduce the
accuracy from $\approx 95\%$ to $\approx 89\%$ on demand to reclaim $\approx 3\times$ EDP.
Similar trends could be seen in all other datasets.

Figure \ref{fig:contour-grove} shows the behavior described earlier. For example, looking at figure
\ref{fig:cont-mnist}, we notice that just by changing the confidence threshold, we can
achieve up to $\approx 10\times$ energy dissipation while having a minimal accuracy penalty.
Note that the numbers presented are energy dissipation, and not EDP. It is fair mentioning
that the latency of the FoG implementation depends on the computation uncertainty. For
example, if uncertainty is very low, only a single grove will be used for computation.
However, in the worst case scenario, uncertainty would be very high, and the computation
would continue up until all groves have tried the input.
}
\begin{table}[tpbh]
  \centering
  \begin{small}
  \begin{tabular}{lccccccc}
    \toprule
    ~                    & \multicolumn{2}{c}{SVM}      &         & &       & \multicolumn{2}{c}{FoG}     \\
    Dataset                     & ${lr}$      & ${rbf}$        & MLP     & CNN   & RF      & $max$      & $opt$  \\
    \midrule
    \texttt{ISOLET} & 69            & 93           & 87      & 94    & 92      & 91   & 90     \\ 
    \texttt{Penbase} & 86            & 95           & 91      & 96    & 96      & 93   & 93     \\
    \texttt{MNIST} & 82            & 95           & 87      & 96    & 96      & 94   & 93     \\
    \texttt{Letter} & 78            & 93           & 93      & 96    & 95      & 85   & 85     \\
    \texttt{Segment.}     & 67            & 91           & 91      & 96    & 95      & 94   & 92 \\
    \midrule
    \texttt{ISOLET} & 5.9       & 980      & 82.5      & 1150    & 41     & 49   & 30 \\ 
    \texttt{Penbase} & 0.4       & 18       & 13.3      & 186     & 16     & 14   & 7.1 \\
    \texttt{MNIST} & 6.1       & 1020      & 93        & 1300    & 43    & 47   & 38 \\
    \texttt{Letter} & 0.5       & 19       & 13.7      & 192     & 16     & 12.9  & 7.6 \\
    \texttt{Segment.} & 0.6       & 26       & 14.5      & 203     & 13     & 9   & 4.7 \\
    \midrule
    Area       & 0.13  & 0.53  & 0.93 & 2.1 & 1.38 & 1.9 & 1.9\\
    \bottomrule
  \end{tabular}
  \caption{Accuracy (top) and Energy dissipation (bottom) in nJ per classification for different datasets~\cite{uci}. Frequency is fixed at 1 $GHz$ for all datasets. SVM$_{lr}$ and SVM$_{rbf}$ show the results for SVM with linear and RBF kernels; FoG$_{max}$ and FoG$_{opt}$ show the results for FoG with its threshold set to maximum and to the optimal accuracy tuning point, respectively. The area results are in mm$^2$}.
  \label{table:datasets}
  \end{small}
\end{table}

%% file: conclusion.tex
\section{Conclusion}
\label{sec:conclusion}
In this work we have compared the lightweight RF classification
algorithm with heavyweight classification algorithms like\ignore{Convolutional Neural
Networks (CNNs), Multi-Layer Perceptron (MLP) Networks and Support Vector
Machines (SVMs)} CNN, MLP, and SVM in terms of accuracy and energy efficiency. We proposed a novel FoGs approach to RF implementation that can
dynamically trade-off accuracy for energy efficiency at run time, while achieving
accuracy comparable to traditional RFs. The proposed FoG approach
examines decision confidence for each input, and allocates the computational
resources depending on the input's uncertainty levels. We implemented the FoG
using a 40 nm technology, and tested it using the datasets provided by the UCI
repository. The evaluation results show that the accuracy of the traditional RF
classifier is comparable (if not larger) to CNN for all datasets that we
considered and at the same time RF consumes $\approx$15$\times$, $\approx$1.7$\times$, and $\approx$23$\times$
less energy than SVM$_{RBF}$, MLP and CNN, respectively. The maximum achievable
accuracy of the FoG implementation is lower than RF and CNN by 3.2\% and 4\%,
respectively, but FoG classifiers have $\approx$1.7$\times$ and $\approx$35$\times$ lower energy than RF and
CNN, respectively.
\ignore{All these numbers have to be changed to the ones in the table!}
\ignore{All the references need to be trimmed to 2 lines maybe 3 lines.}